# StudDFT: A quantum chemistry program for educational and research purposes

Stanislav K. Ignatov
Department of Chemistry, Lobachevsky State University of Nizhny Novgorod,
Nizhny Novgorod, Russia
E-mail: skignatov@gmail.com

**Abstract**

StudDFT is a quantum chemistry program for Hartree–Fock (HF), density functional theory (DFT), and second-order Møller–Plesset (MP2) calculations of small and medium-sized molecules. It is intended primarily for teaching and student research, but its accuracy suffices for routine research calculations. Version 1.30.2 implements the RHF/UHF, RKS/UKS, and RMP2/UMP2/PUMP2 schemes in Gaussian basis sets, including effective core potentials (ECP); LDA, GGA (BLYP, BP86, BPW91, PBE), meta-GGA ($r^2$SCAN), global hybrid (B3LYP, PBE0) and range-separated hybrid ($\omega$B97, $\omega$B97X, $\omega$B97X-D) functionals; analytical gradients and Hessians; geometry optimization, transition-state and reaction-path (IRC) searches, potential energy surface scans, harmonic frequencies and IR intensities, thermochemistry, counterpoise corrections, D3(BJ) dispersion, polarizable continuum solvation, density fitting (RI-J/RI-JK), linear-response TDDFT, and fractional-occupation smearing. The program has been verified against ORCA and PySCF on 50 molecules (30 for optimizations): the mean absolute deviations of the single-point energies do not exceed 7 μHa for HF, MP2, and the LDA, GGA, and global hybrid functionals except PBE and PBE0, whose systematic shifts of several tens of μHa are traced mainly to the value of $\mu$ in the PBE exchange; those of the optimized interatomic distances and harmonic frequencies do not exceed $1.2\cdot10^{-4}$ Å and 1.2 cm$^{-1}$. Corrections made since the first public version (0.82), mainly affecting vibrational frequencies, are documented with their effect on the verification results. Unlike large multi-language packages, StudDFT is a monolithic Fortran code with OpenMP parallelization, which simplifies its use in courses. Binaries for Windows and Linux, the documentation, and all test inputs and outputs are freely available.

## 1 Introduction

The progress of quantum chemistry is inseparably linked to the development of software, and the limits of applicability of a particular theoretical method often depend on the specific implementation of the program. On the other hand, the class of problems to be solved dictates the need to develop software that implements the quantum chemical methods best suited for them.

A special place among these methods is occupied by density functional theory (DFT). As a compromise between accuracy and computational cost, DFT has become the *de facto* standard for the calculation of medium-sized molecules, as confirmed by its presence in the overwhelming majority of modern software packages. From an educational standpoint, this theory is especially convenient, since in a relatively short time and without sophisticated computing equipment it allows one to obtain results that correlate directly with experiment. By now, a well-developed ecosystem of programs supporting DFT calculations has taken shape. These programs can be tentatively divided into three categories according to their main purpose. First, there are the commercial packages Gaussian [1], Q-Chem [2], Turbomole [3], Molcas [4], Molpro [5] and others, which are the result of many years of work by large teams of developers and provide the most complete set of methods, functionals, and basis sets. These programs have undergone repeated optimization and validation and are regarded as reference programs in terms of both reliability and performance. However, their use is restricted by commercial licenses, and their installation and use often require considerable expertise. Second, over the past two decades several large projects have emerged whose goal is to provide the research community with freely available quantum chemistry programs. The most notable of these are ORCA [6], NWChem [7], GAMESS [8], Psi4 [9], and PySCF [10]. Unfortunately, the level of training of junior chemistry students, as well as the limited access to high-performance computing hardware (including Linux systems), often makes it impossible to install and use these codes on their own computers, because of the high hardware requirements, the need to compile the source code, and the need to install additional libraries.

A separate category comprises codes written specifically for educational purposes. These include the Crawford group's programming projects [11], with sequential implementations of HF and post-HF methods in the form of exercises; the Psi4NumPy project [12], an environment of readable reference implementations and interactive teaching modules based on Python/NumPy; the education-oriented PyQInt code [13], which implements HF with Gaussian integrals, SCF, and geometry optimization; and the early teaching package PyQuante [14]. The strength of these codes is their educational focus. Their weakness lies in limited functionality: all of them are oriented toward HF and post-HF methods, and full-fledged DFT with analytical derivatives is essentially absent among them.

The landscape described leaves a notable methodological gap. On the one hand, a simple teaching code is usually unable to calculate a real medium-sized molecule in a reasonable time on a student's laptop. On the other hand, a research package capable of such a calculation usually requires a large amount of disk space, self-compilation, and the installation of additional libraries, so that its installation and mastering become a separate, labor-intensive process.

The aim of this work is to develop a program that occupies an intermediate position in this spectrum. StudDFT is a compact quantum chemistry package implementing the basic methods of quantum chemistry (HF, DFT, and MP2 as a means of additional refinement and verification of the results). It enables geometry optimization and the calculation of vibrational frequencies, electronic properties, and thermodynamic functions for a broad class of compounds.

The principles that guided the development of the program were as follows. The program should cover the quantum chemical theories with which undergraduate, master's, and doctoral students can quickly carry out educational and research calculations on their personal computers (HF, DFT, MP2); use a wide range of basis sets with the possibility of extending them; be oriented toward a broad class of small and medium-sized molecular systems; have an interface similar to those of modern professional programs; and allow the simplest possible installation and use on computers and laptops with limited memory and disk space. To accomplish educational tasks, the program should be able to perform geometry optimization, PES scanning, calculation of vibrational frequencies, transition-state searches, and evaluation of the basis-set superposition error and solvent effects.

Thanks to the development of artificial intelligence systems, the development of programs of this level is today being radically simplified and accelerated. The developer, however, is required to have a good understanding of modern algorithms of computational quantum chemistry, their capabilities, and how they fit together, taking into account the trade-offs between complexity, accuracy, and performance. A special concern when using a new development technology is the comprehensive verification of the correctness of the code. This verification is based on thorough testing of the program against reference professional codes.

The first description of StudDFT, corresponding to program version 0.82, has been accepted for publication in Zhurnal Fizicheskoi Khimii (in Russian) [15]. The present paper is an extended English version that describes version 1.30.2, which is available for Windows and Linux, and repeats the verification with the corrected program. The theoretical background is summarized in Section 2, the numerical algorithms in Section 3, and the program capabilities in Section 4. Section 5 presents the verification of the program against reference codes, including an analysis of the corrections made after version 0.82. Additional data are given in the Supplementary Information (Appendix).

# 2 Theory

Below we briefly present the theoretical concepts underlying the methods implemented in StudDFT. A detailed description of the computational formulas can be found in the monographs [16–18] and the references therein.

## 2.1 Hartree–Fock and Kohn–Sham equations

Within the Hartree–Fock (HF) method, the one-electron molecular orbitals $\varphi_p$ are sought as linear combinations of the basis functions $\chi_\mu$:

$$\varphi_p = \sum_{\mu} C_{\mu p}\, \chi_\mu.$$

The Hartree–Fock–Roothaan equations are written in matrix form as

$$\mathbf{FC} = \mathbf{SC}\boldsymbol{\varepsilon},$$

where **F** is the Fock matrix (Fockian), **S** is the overlap matrix of the basis functions, **C** is the matrix of orbital coefficients, and **ε** is the diagonal matrix of orbital energies. The matrix **P** with the elements

$$P_{\mu\nu} = \sum_{i \in \mathrm{occ}} C_{\mu i}\, C_{\nu i}$$

is called the density matrix (in StudDFT it is defined without the factor of 2). The sum runs over the $n_{\mathrm{occ}} = n_{\mathrm{elec}}/2$ occupied spatial orbitals; for the unrestricted formulation, the density matrices $\mathbf{P}^\alpha$ and $\mathbf{P}^\beta$ are defined separately.

Within Kohn–Sham (KS) theory [19], the system is described by equations that have the same matrix form as in the HF method, but with the Hamiltonian (hereinafter, by analogy with the HF method, called the Fockian) augmented by an exchange-correlation contribution. The Fockian unifying both HF and DFT is written for a closed-shell system as

$$\mathbf{F} = \mathbf{H}^{\mathrm{core}} + 2\,\mathbf{J}(\mathbf{P}) - c_{\mathrm{HF}}\,\mathbf{K}(\mathbf{P}) + \mathbf{V}^{xc}(\mathbf{P}),$$

where $\mathbf{H}^{\mathrm{core}}$ is the one-electron core Hamiltonian, which includes the electron kinetic energy, the nuclear attraction, and the ECP contribution when present,

$$\mathbf{H}^{\mathrm{core}} = \mathbf{T} + \mathbf{V}^{\mathrm{nuc}} + \mathbf{V}^{\mathrm{ECP}},$$

and $\mathbf{J}(\mathbf{P})$ and $\mathbf{K}(\mathbf{P})$ are the Coulomb and exchange matrices,

$$J_{\mu\nu}(\mathbf{P}) = \sum_{\lambda\sigma} P_{\lambda\sigma}\,(\mu\nu|\lambda\sigma), \qquad K_{\mu\nu}(\mathbf{P}) = \sum_{\lambda\sigma} P_{\lambda\sigma}\,(\mu\lambda|\nu\sigma),$$

$$(\mu\nu|\lambda\sigma) = \iint \chi_\mu(\mathbf{r}_1)\chi_\nu(\mathbf{r}_1)\,\frac{1}{r_{12}}\,\chi_\lambda(\mathbf{r}_2)\chi_\sigma(\mathbf{r}_2)\,d\mathbf{r}_1\,d\mathbf{r}_2.$$

Here $c_{\mathrm{HF}}$ is the fraction of exact (HF) exchange ($c_{\mathrm{HF}} = 1$ for pure HF, 0 for LDA, GGA and meta-GGA functionals, 0.20 for B3LYP, and 0.25 for PBE0), and $\mathbf{V}^{xc}(\mathbf{P})$ is the exchange-correlation potential (zero for pure HF), evaluated by numerical integration on a grid:

$$V_{\mu\nu}^{xc} = \int \chi_\mu(\mathbf{r})\, v_{xc}\big(\rho(\mathbf{r}), \nabla\rho(\mathbf{r})\big)\, \chi_\nu(\mathbf{r})\, d\mathbf{r}$$

(for GGA and meta-GGA functionals, $v_{xc}$ is understood as the corresponding differential operator). The electronic energy of the system is

$$E_{\mathrm{elec}} = \mathrm{Tr}[\mathbf{P}(\mathbf{H}^{\mathrm{core}} + \mathbf{F})] + E_{xc} - \mathrm{Tr}[\mathbf{P}\mathbf{V}^{xc}]$$

(the last two terms vanish for HF), where the exchange-correlation energy is

$$E_{xc} = \int \varepsilon_{xc}[\rho(\mathbf{r})]\, \rho(\mathbf{r})\, d\mathbf{r}$$

and $\varepsilon_{xc}$ is the exchange-correlation energy per particle. The total energy of the system is the sum of the electronic energy and the internuclear repulsion energy:

$$E_{\mathrm{tot}} = E_{\mathrm{elec}} + E_{\mathrm{nuc}} = E_{\mathrm{elec}} + \sum_{A<B} \frac{Z_A Z_B}{R_{AB}}.$$

The spin-restricted formulation (closed shell, RHF/RKS) is applied to singlet systems with an even number of electrons: the $\alpha$ and $\beta$ orbitals are taken to be pairwise identical. In the spin-unrestricted formulation (UHF/UKS) [20], the $\alpha$ and $\beta$ orbitals are varied independently, which leads to two sets of Roothaan–Hall equations,

$$\mathbf{F}^{\alpha}\mathbf{C}^{\alpha} = \mathbf{S}\mathbf{C}^{\alpha}\boldsymbol{\varepsilon}^{\alpha}, \qquad \mathbf{F}^{\beta}\mathbf{C}^{\beta} = \mathbf{S}\mathbf{C}^{\beta}\boldsymbol{\varepsilon}^{\beta},$$

with separately defined Fock matrices

$$\mathbf{F}^{\sigma} = \mathbf{H}^{\mathrm{core}} + \mathbf{J}(\mathbf{Q}) - c_{\mathrm{HF}}\,\mathbf{K}(\mathbf{P}^{\sigma}) + \mathbf{V}^{xc,\sigma}, \qquad \sigma = \alpha, \beta,$$

where $\mathbf{Q} = \mathbf{P}^{\alpha} + \mathbf{P}^{\beta}$ is the total density matrix and $\mathbf{V}^{xc,\alpha}$, $\mathbf{V}^{xc,\beta}$ are the spin-dependent exchange-correlation potentials.

The expectation value of the squared spin operator in UHF is

$$\langle \hat{S}^2 \rangle = S_z(S_z + 1) + N_{\beta} - \sum_{ij} |\langle i_{\alpha} | j_{\beta} \rangle|^2,$$

where $S_z = (N_{\alpha} - N_{\beta})/2$, and the summation is carried out over the occupied molecular orbitals *i* and *j*. For a pure spin state, $\langle \hat{S}^2 \rangle = S(S+1)$; a deviation indicates admixtures of higher spin states (spin contamination).

## 2.2 Analytical gradient and Hessian

The gradient of the total energy with respect to the coordinates of nucleus *A* is written in StudDFT as

$$\frac{dE_{\mathrm{tot}}}{d\mathbf{R}_A} = \mathbf{g}_A^{\mathrm{nuc}} + \mathbf{g}_A^{1e} + \mathbf{g}_A^{2e} + \mathbf{g}_A^{\mathrm{ECP}} + \mathbf{g}_A^{xc},$$

$$\mathbf{g}_A^{\mathrm{nuc}} = \frac{\partial E_{\mathrm{nuc}}}{\partial \mathbf{R}_A}, \qquad \mathbf{g}_A^{1e} = -\mathrm{Tr}\left[\mathbf{D}\frac{\partial \mathbf{S}}{\partial \mathbf{R}_A}\right] + \mathrm{Tr}\left[\mathbf{Q}\frac{\partial \mathbf{T}}{\partial \mathbf{R}_A}\right] + \mathrm{Tr}\left[\mathbf{Q}\frac{\partial \mathbf{V}^{\mathrm{nuc}}}{\partial \mathbf{R}_A}\right],$$

$$\mathbf{g}_A^{2e} = \frac{1}{2}\sum_{\sigma=\alpha,\beta} \mathrm{Tr}\,[\mathbf{P}^{\sigma}\,\mathbf{G}^{\sigma,A}], \qquad \mathbf{g}_A^{\mathrm{ECP}} = \mathrm{Tr}\left[\mathbf{Q}\frac{\partial \mathbf{V}^{\mathrm{ECP}}}{\partial \mathbf{R}_A}\right], \qquad \mathbf{g}_A^{xc} = \frac{\partial E_{xc}}{\partial \mathbf{R}_A}\Big|_{\mathbf{P}}.$$

Here $\mathbf{D} = \mathbf{W}^{\alpha} + \mathbf{W}^{\beta}$ is the energy-weighted density matrix, $W_{\mu\nu}^{\sigma} = \sum_{i\in\mathrm{occ}} \varepsilon_i^{\sigma}\, C_{\mu i}^{\sigma} C_{\nu i}^{\sigma}$, and $\mathbf{G}^{\sigma,A} = \mathbf{J}^{A}(\mathbf{Q}) - c_{\mathrm{HF}}\mathbf{K}^{A}(\mathbf{P}^{\sigma})$ is the two-electron part of the Fockian built from the derivative integrals $\partial(\mu\nu|\lambda\sigma)/\,\partial\mathbf{R}_A$ (for a closed-shell system, $\mathbf{Q} = 2\mathbf{P}$, and $\mathbf{P}^{\alpha} = \mathbf{P}^{\beta} = \mathbf{P}$). The derivatives of the one-electron and ECP matrices include both the derivatives of the basis functions and, for $\mathbf{V}^{\mathrm{nuc}}$ and $\mathbf{V}^{\mathrm{ECP}}$, the derivatives of the operators themselves.

The Hessian is written as

$$H_{xy} = \frac{\partial^2 E_{\text{tot}}}{\partial x\, \partial y} = H_{xy}^{\text{nuc}} + H_{xy}^{1e} + H_{xy}^{2e} + H_{xy}^{xc,\text{func}} + H_{xy}^{xc,\text{weight}} + H_{xy}^{\text{resp}}.$$

The first five terms are the second derivatives, with respect to the Cartesian coordinates *x* and *y*, of the nuclear repulsion energy, of the one-electron part (including the ECP, if present), of the two-electron part, of the exchange-correlation energy at fixed grid weights, and the contribution of the derivatives of the grid weights, respectively. These contributions are computed at a fixed density matrix. The last term accounts for the response of the electron density to the displacement of the nuclei and is computed by the coupled-perturbed SCF method (CPHF/CPKS). In version 1.30.2, the contributions of the first derivatives of the Becke weights are included (also in the right-hand sides of the CPKS equations), whereas the term with the second derivatives of the weights, $\sum_g \left(\partial^2 w_g / \partial x\, \partial y\right) f_g$ ($f_g$ is the exchange-correlation energy density at the grid point *g*), is omitted. The omitted term is small: in our tests, it changes the frequencies of covalently bound molecules by about 0.1 cm$^{-1}$ (e.g., the bending modes of $SiH_4$; Section 5.3), but it can reach a few cm$^{-1}$ for soft intermolecular modes (4.6 cm$^{-1}$ for a libration of 47 cm$^{-1}$ in $(H_2O)_2$); the program prints a corresponding note. The Hessian in the MP2/UMP2/PUMP2 methods is formed by numerical differentiation (central differences) of the analytical gradients.

## 2.3 Exchange-correlation functionals

StudDFT implements exchange and correlation functionals of several types, classified according to the well-known "Jacob's ladder" [21].

*Local density approximation (LDA).* This approximation is implemented in two variants, SVWN3 and SVWN5 (the latter is used by default and is hereafter denoted SVWN). Both functionals depend only on the local density $\rho(\mathbf{r})$ and combine the Dirac exchange functional [22],

$$\varepsilon_x^{\text{LDA}}(\rho) = -\frac{3}{4}\left(\frac{3}{\pi}\right)^{1/3}\rho^{1/3},$$

with the Vosko–Wilk–Nusair correlation functional VWN3 or VWN5 [23], which is an analytical function of $r_s = (3/(4\pi\rho))^{1/3}$. The choice between these two variants makes it possible to reproduce the results of the widely used quantum chemistry programs, each of which implements one of them.

*Generalized gradient approximation (GGA).* In this approximation, the exchange-correlation functional depends both on $\rho$ and on the squared density gradient $\sigma = |\nabla\rho|^2$. The general form of the GGA exchange is written through the dimensionless density gradient

$$s = \frac{|\nabla\rho|}{2(3\pi^2)^{1/3}\rho^{4/3}}$$

and the enhancement factor $F_x(s)$:

$$\varepsilon_x^{\text{GGA}}(\rho, s) = \varepsilon_x^{\text{LDA}}(\rho)\, F_x(s).$$

StudDFT implements the B88 [24] and PBE [25] exchange functionals with the corresponding parametrizations of $F_x(s)$. The GGA correlation is implemented in four variants: LYP [26], a correlation functional constructed without an explicit LDA limit; the Perdew–Wang PW91 functional [27, 28]; the PBE correlation [25]; and P86 [29], an early gradient correction to the LDA. The combinations implemented in StudDFT are: BLYP = B88 + LYP; BP86 = B88 + P86; BPW91 = B88 + PW91; PBE = PBE exchange + PBE correlation. For the local part of the P86 correlation, the Perdew–Wang PW92 parametrization [30] is used, as in ORCA. ORCA, however, uses the parameters *A* = 0.0310907, 0.01554535, and 0.0168869 of the reference code of Perdew and Wang (labelled PW91-LDA in its output; libxc uses the same values in its PBE and r$^2$SCAN correlation), whereas StudDFT uses the published values *A* = 0.031091, 0.015545, and 0.016887 [30] in all PW92-based correlation functionals (P86, PW91, PBE, r$^2$SCAN, and the ωB97 family); this difference lowers the StudDFT energies of the molecules of Table 3 by 3–4 μHa on average (r$^2$SCAN: 1.4 μHa; Section 5.2). Note that the BP86 functional of Gaussian and libxc uses the Perdew–Zunger PZ81 parametrization [31] instead; with identical quadratures, the resulting BP86 energies differ from those of StudDFT by 75 μHa for $CH_4$ and by 5.0 mHa for $O_3$. In the PBE exchange, the constants $\kappa = 0.804$ and $\mu = 0.21951$ of the original publication [25] are used, and in the PBE correlation the values $\beta = 0.066725$ and $\gamma = 0.031091$ given there (see Section 5.1 for the consequences of these choices in comparisons with ORCA and libxc [32]).

*Meta-GGA.* The r$^2$SCAN functional [33], a regularized and restored version of SCAN, depends additionally on the kinetic-energy density $\tau(\mathbf{r}) = \frac{1}{2}\sum_i |\nabla\varphi_i(\mathbf{r})|^2$ (the sum runs over the occupied spin orbitals; the spin components $\tau_\alpha$ and $\tau_\beta$ are used for open shells). Its contribution to the Fockian contains the term $\frac{1}{2}\int (\partial f_{xc}/\,\partial\tau)\,\nabla\chi_\mu \cdot \nabla\chi_\nu\, d\mathbf{r}$.

*Hybrid functionals.* Global hybrid functionals include a fraction of exact exchange combined with GGA exchange and correlation. The implemented variants are B3LYP [34, 35] and PBE0 [36]:

$$E_{xc}^{\mathrm{B3LYP}} = 0.20\,E_x^{\mathrm{HF}} + 0.08\,E_x^{\mathrm{LDA}} + 0.72\,E_x^{\mathrm{B88}} + 0.81\,E_c^{\mathrm{LYP}} + 0.19\,E_c^{\mathrm{VWN5}},$$
$$E_{xc}^{\mathrm{PBE0}} = 0.25\,E_x^{\mathrm{HF}} + 0.75\,E_x^{\mathrm{PBE}} + E_c^{\mathrm{PBE}}.$$

The definition of B3LYP with VWN5 in the correlation part corresponds to the formulation discussed in Ref. [37] and is consistent with the ORCA program. The Gaussian program uses VWN3 in the same role, which leads to a small but systematic difference in the energies; to reproduce Gaussian results, the user can select the B3LYP3 variant. The PBE0 formulation corresponds to the PBE1PBE functional of Gaussian.

*Range-separated hybrids.* In the $\omega$B97 family [38, 39], the Coulomb operator is split by the error function, $1/r_{12} = \mathrm{erf}(\omega r_{12})/r_{12} + \mathrm{erfc}(\omega r_{12})/r_{12}$; the long-range part of the exchange is treated exactly, and the short-range part by a modified B97-type functional. StudDFT implements $\omega$B97 ($\omega = 0.4$ Bohr$^{-1}$, no short-range HF exchange), $\omega$B97X ($\omega = 0.3$ Bohr$^{-1}$, 15.77% of short-range HF exchange), and $\omega$B97X-D ($\omega = 0.2$ Bohr$^{-1}$, 22.2% of short-range HF exchange, with the Chai–Head-Gordon damped $C_6/R^6$ dispersion correction [39]); the parameter $\omega$ can be changed by the user ("$\omega$-tuning"). Since the integrals over the attenuated operator differ from the ordinary ones only in the argument and the prefactor of the Boys function [40], no new recursions are required.

All implemented functionals have a spin-dependent formulation in which

$$\varepsilon_{xc} = \varepsilon_x(\rho_\alpha, \rho_\beta) + \varepsilon_c(\rho_\alpha, \rho_\beta, \sigma_{\alpha\alpha}, \sigma_{\alpha\beta}, \sigma_{\beta\beta}),$$

where $\sigma_{\alpha\beta} = \nabla\rho_\alpha \cdot \nabla\rho_\beta$. For the exchange, the spin-scaling relation is applied [41]; for the correlation functionals, the corresponding formulas are more complex and are given in the original publications. To estimate the spin contamination, the program by default computes $\langle \hat{S}^2 \rangle$ for all open-shell systems.

All first and most second derivatives of the functionals are implemented analytically, which is important for the analytical Hessian and the coupled-perturbed equations (CPHF/CPKS). Nevertheless, numerical implementations (central finite differences with respect to $\rho$ and $\sigma$ with a relative step of $10^{-5}$) are retained in the code for all functionals as a verification mechanism; the analytical and numerical derivatives agree to about $10^{-9}$ in relative units on the test cases.

## 2.4 The MP2 method and spin-projected PUMP2

Second-order Møller–Plesset perturbation theory (MP2) [42] gives a second-order correction to the HF energy which, in the closed-shell case (RMP2), has the form

$$E_{\mathrm{MP2}} = \sum_{ij}^{\mathrm{occ}} \sum_{ab}^{\mathrm{vir}} \frac{(ia|jb)\,[2(ia|jb) - (ib|ja)]}{\varepsilon_i + \varepsilon_j - \varepsilon_a - \varepsilon_b},$$

where $i, j$ run over the occupied and $a, b$ over the virtual canonical HF orbitals, $\varepsilon_p$ are the orbital energies, and the integrals in the MO basis are obtained from the AO integrals by a four-index transformation. In the UMP2 method, the formula is generalized to the spin-orbital basis with separate summation over the $\alpha\alpha$, $\alpha\beta$, and $\beta\beta$ components.

For UMP2, spin contamination is especially pronounced and may lead to significant unphysical contributions to the energy. The PUMP2 method [43] removes the largest part of these admixtures by annihilating the contribution of the next-higher spin state with the operator

$$\hat{A}_{S+1} = \frac{\hat{S}^2 - (S+1)(S+2)}{S(S+1) - (S+1)(S+2)}$$

applied to the UHF wave function. At the MP2 level, this reduces to an energy correction, $E_{\mathrm{PUMP2}} = E_{\mathrm{UMP2}} + \Delta E_{\mathrm{PUMP2}}$, where $\Delta E_{\mathrm{PUMP2}}$ is expressed through the overlaps between the UHF wave function and the function obtained after annihilation [43]. In StudDFT, the RMP2/UMP2/PUMP2 methods are available for energy calculations and for geometry optimization (with an analytical gradient for RMP2/UMP2 and a numerical one for PUMP2).

### 2.5 Excited states and solvation

*Linear-response TDDFT.* Vertical excitation energies $\omega_n$ and oscillator strengths are obtained from the Casida equations [44]

$$\begin{pmatrix} \mathbf{A} & \mathbf{B} \\ \mathbf{B} & \mathbf{A} \end{pmatrix} \begin{pmatrix} \mathbf{X}_n \\ \mathbf{Y}_n \end{pmatrix} = \omega_n \begin{pmatrix} \mathbf{1} & \mathbf{0} \\ \mathbf{0} & -\mathbf{1} \end{pmatrix} \begin{pmatrix} \mathbf{X}_n \\ \mathbf{Y}_n \end{pmatrix},$$

where the matrices **A** and **B** contain the orbital-energy differences, the Coulomb and (scaled) exchange integrals, and the exchange-correlation kernel. In the Tamm–Dancoff approximation (TDA) [45], $\mathbf{B} = \mathbf{0}$. The equations are solved iteratively in a subspace for singlet and/or triplet excitations of closed-shell (RHF/RKS) references with LDA, GGA, global and range-separated hybrid functionals (TDHF and CIS for $c_{\mathrm{HF}} = 1$).

*Polarizable continuum.* The solvent is described by apparent surface charges on a molecular cavity formed by interlocking atomic spheres (Bondi radii [46] scaled by 1.2). The cavity surface is discretized with atom-centred Lebedev grids, and a smooth switching function is applied to the surface elements near sphere intersections. The IEFPCM [47] (default), C-PCM [48], and COSMO [49] formulations are implemented, together with a linear-response PCM kernel for TDDFT in the equilibrium and nonequilibrium regimes.

## 3 Numerical algorithms

In developing the program, we have tried to use the most modern schemes described in the publicly available literature. A deliberate design decision in StudDFT is the selection of algorithms optimal for small and medium-sized systems (up to about 500 basis functions): for this range, recursive schemes balanced between implementation complexity and efficiency have been chosen. Larger systems can also be calculated without any changes to the program, at the cost of lower performance. All analytical integrals are evaluated over Cartesian Gaussian functions, regardless of whether a Cartesian or a spherical atomic basis is used.

### 3.1 Basis sets

All calculations are performed in Gaussian basis sets. Cartesian (default) or pure spherical basis functions can be used. The maximum angular momentum of the basis functions is $L_{\max} = 4$ ($g$ functions), which is sufficient for most practical DFT and MP2 calculations. Segmented contractions are used; general contraction is not implemented, because for HF, DFT, and MP2 calculations of small and medium-sized molecules it does not lead to a noticeable speed-up but complicates the program code. The basis-set library is an open directory with files describing the basis sets and ECPs in a format compatible with the Basis Set Exchange (BSE) [50]; the user can therefore add a basis set by downloading the corresponding file from BSE. The built-in library contains the Pople sets from STO-3G to 6-311++G(3df,2pd), the correlation-consistent sets (aug-)cc-pV*n*Z (*n* = D, T, Q), and LANL2DZ. Composite basis sets can be used by assigning basis sets (and ECPs) to individual atoms or elements.

### 3.2 One-electron integrals

One-electron integrals in StudDFT are computed by two different recursive schemes: the overlap integrals **S** and the kinetic-energy integrals **T** by the Obara–Saika scheme [51], and the nuclear-attraction integrals $\mathbf{V}^{\text{nuc}}$ by the McMurchie–Davidson scheme [52]. The Obara–Saika method is the shortest path to **S** and **T**: these integrals are products of one-dimensional integrals along the Cartesian axes, and the angular-momentum-raising recursion along each axis has a concise form with a small number of terms. The McMurchie–Davidson method, by contrast, expands the product of two Cartesian Gaussians in Hermite Gaussians $\Lambda_t$ using the so-called $E$ coefficients,

$$(x - A_x)^{l_a}(x - B_x)^{l_b} e^{-p(x-P_x)^2} = \sum_{t=0}^{l_a+l_b} E_t^{l_a l_b}\, \Lambda_t(x; p, P_x),$$

after which the integral is reduced to a family of integrals expressed through the Boys function. This route is better suited for $\mathbf{V}^{\text{nuc}}$.

### 3.3 The Boys function

The Boys function

$$F_m(T) = \int_0^1 t^{2m}\, e^{-Tt^2}\, dt$$

is the "seed function" of all two-electron integrals with the Coulomb operator, as well as of $\mathbf{V}^{\text{nuc}}$ and a large fraction of the ECP integrals, so the accuracy of its evaluation determines the accuracy of all subsequent calculations. StudDFT implements two independent schemes for $F_m(T)$: a fast interpolation based on precomputed tables of node values (the default), and a more accurate iterative scheme of Cook [53], which can be requested with a special keyword. For Hessian calculations, the program automatically switches to the Cook scheme as a precaution against the amplification of small interpolation errors by numerical differentiation.

### 3.4 Two-electron integrals

The two-electron integrals $(\mu\nu|\lambda\sigma)$ are computed by the Head-Gordon–Pople (HGP) method [54]: a vertical recursion (VRR) over the four centres, followed by a horizontal recursion (HRR) that transfers angular momentum between the centres of each pair; the HRR route is chosen automatically for each shell quartet. The choice of HGP over the alternatives (the McMurchie–Davidson, Obara–Saika, and Rys methods) is dictated by the balance between implementation complexity and performance. HGP has an additional practical advantage: the VRR phase is computed once for the *bra* and *ket* pairs, while HRR is a comparatively cheap algebraic operation, which makes it possible to use caching efficiently for shells of moderate size.

To screen out small integrals, StudDFT uses the Schwarz inequality

$$|(\mu\nu|\lambda\sigma)| \le \sqrt{(\mu\nu|\mu\nu)}\,\sqrt{(\lambda\sigma|\lambda\sigma)}.$$

At the initialization stage, the program computes and stores the diagonal values for all shell pairs; the list of shell pairs whose diagonal exceeds a given threshold ($10^{-12}$ Hartree by default) is formed once and used in all SCF iterations.

*Density fitting.* As an option, the Coulomb (RI-J) or the Coulomb and exchange (RI-JK) matrices are built with the resolution of the identity, $(\mu\nu|\lambda\sigma) \approx \sum_{PQ} (\mu\nu|P)\ [\mathbf{V}^{-1}]_{PQ}\ (Q|\lambda\sigma)$, where $P, Q$ are auxiliary functions and $V_{PQ} = (P|Q)$ [55]. The def2-universal JFIT [56] (H–Ar, extended by decontracted sets for K–Rn) and JKFIT [57] (H–Ar) auxiliary bases are loaded automatically. For pure functionals with RI-J, the four-index integral phase is skipped entirely. Analytical RI gradients and an analytical RI-J Hessian (pure functionals, closed shells) are available.

## 3.5 Integrals of effective core potentials

ECPs are supported in the form of Hay–Wadt-type potentials (LANL2DZ) [58–60],

$$\hat{V}^{\mathrm{ECP}} = U_{\mathrm{loc}}(r) + \sum_{l} \left(U_l(r) - U_{\mathrm{loc}}(r)\right) \hat{P}_l,$$

where $\hat{P}_l$ is the projector onto angular momentum *l* about the ECP centre, and $U_l(r)$ are radial functions defined as sums of Gaussians multiplied by integer powers of *r*. The matrix elements of $\hat{V}^{\mathrm{ECP}}$ reduce to two types of integrals: "local" (type 1) integrals of $U_{\mathrm{loc}}(r)$ with two Gaussian functions, and "semilocal" (type 2) integrals containing the projector $\hat{P}_l$.

Both types of integrals are implemented in two independent ways. In the method called "analytical" in the program, the local integrals are computed by the McMurchie–Davidson scheme [61] (modified *R* integrals with the absorbed ECP Gaussian), and the semilocal ones by the half-numerical angular-projection method of Flores-Moreno et al. [62], in which the angular integration is analytical. In the numerical method, integrals of both types are computed by direct quadrature centred on the ECP atom: a 50-point Gauss–Legendre radial rule on an interval adapted to the exponents involved, combined with an 18×36 product angular grid (Gauss–Legendre in $\cos\theta$ and uniform in $\phi$), i.e., 32 400 points. This method is simple but requires many quadrature points. The analytical method is about 10 times faster; however, for some systems (e.g., $AlF_3$ and iodine compounds) its accuracy is noticeably worse (Section 5.1), and the numerical method is therefore the default for the energy. The ECP gradient is evaluated consistently with the chosen method: in the numerical method, the analytically differentiated basis functions are integrated with the same quadrature points (in version 1.30.2, however, with a narrower radial interval than for the energy; see Section 5.1), and in the analytical method, analytical derivative integrals are used. The ECP skeleton contribution to the analytical Hessian is computed in both cases by the quadrature with the second derivatives of the basis functions.

## 3.6 Integration grids for the exchange-correlation functionals

In DFT calculations, the exchange-correlation integrals are evaluated numerically on a molecular grid built as the union of atomic grids whose weights partition space according to the Becke scheme [63]: the polynomial $p(\mu) = \frac{3}{2}\mu - \frac{1}{2}\mu^3$ is iterated three times, and the cell function is $s(\mu) = \frac{1}{2}\left[1 - p\left(p(p(\mu))\right)\right]$. Each atomic grid is the product of a radial and an angular grid. The radial grids of Mura and Knowles (Log3 map) [64], Euler–Maclaurin [65], and Treutler–Ahlrichs M3/M4 [66] are implemented; the angular grids are Lebedev–Laikov grids [67] of standard orders. Twelve predefined grids are selected with the keyword `%grid` (Table S1 of the Supplementary Information); the default is grid 3. In grids 2–7 and 10–12, the angular order is reduced in the inner and outer radial regions (pruning): grids 2–5, 11, and 12 use three tiers defined by the ratio of $r$ to the Bragg–Slater radius, and grids 6, 7, and 10 use five-zone schemes (Table S1). Grids 6–9, 11, and 12 additionally use the atomic size adjustment of the Becke partition with the parameter $\chi$ determined by the Becke method [63] ($\chi$ is the ratio of the atomic radii) or the Treutler–Ahlrichs method [66] ($\chi$ is the square root of this ratio). Grids 11 and 12 are compact but complete (full Lebedev shells, correct weights, correct partitioning) grids intended for the response calculations (CPKS and TDDFT); grid 11 is the default response grid of the analytical Hessian (Section 3.9).

Since version 1.28.0, the Mura–Knowles grids 1–5 use the radial scale parameters $\alpha$ tabulated by these authors [64] (5.0 Bohr, and 7.0 Bohr for the elements of groups 1 and 2; for group 2, since version 1.29.2). In earlier versions, the Bragg–Slater radius was used in the Log3 map by mistake, which truncated the radial grid (e.g., at 4.4 Bohr around O and 2.6 Bohr around H) and caused quadrature errors of up to several mHa with the default grid; the old behaviour can be restored with the keyword `%grid_legacy_mk` only to reproduce earlier results. Grids 6–10, including grid 8 used in all verification calculations of this work, are not affected.

### 3.7 SCF procedure

The SCF procedure for all modes (RHF/UHF/RKS/UKS) is performed by default as a “conventional” SCF with the two-electron integrals stored in RAM. The default buffer size (4 GB) corresponds to calculations with approximately 300–400 basis functions; it can be changed by the user depending on the available memory. If the buffer is insufficient, the program automatically switches to the direct SCF scheme, in which the Fock matrix is built incrementally from the density difference with Schwarz× $|\Delta\mathbf{P}|$ screening and fully rebuilt every 20 iterations. Storing integrals on disk is not supported.

Two initial guesses for the density matrix are available: the eigenvectors of the core Hamiltonian $\mathbf{H}^{\mathrm{core}}$ and the superposition of atomic densities (SAD) [68], which also supports ECP atoms. By default, the core Hamiltonian is used for closed shells, since it is significantly faster, and SAD for open shells, for which the core guess often leads to poor convergence. The density can also be read from a checkpoint file written after every converged SCF, which allows warm restarts and continuation of interrupted calculations.

StudDFT implements several convergence-acceleration techniques: the DIIS method [69], level shifting, damping, and the maximum-overlap method (MOM) [70]. DIIS is used by default with the commutator residual $\mathbf{e} = \mathbf{FPS} - \mathbf{SPF}$ (in the orthogonal basis); the oldest vectors are removed from the DIIS subspace when its condition number becomes too large. For problematic systems (quasi-degenerate systems, transition metals), linear damping of the density,

$$\mathbf{P}_{\mathrm{new}} = (1-\lambda)\,\mathbf{P}_{\mathrm{DIIS}} + \lambda\,\mathbf{P}_{\mathrm{old}},$$

can be applied at the early iterations. A level shift of the virtual orbitals is available through the keyword `%level_shift`; in the automatic mode, a shift of 0.2 Hartree is turned on if convergence has not been reached after 100 iterations. For open shells, the orbital occupation during the first iterations is controlled by MOM, followed by strict Aufbau occupation (alternatively, strict Aufbau or MOM throughout can be requested). For closed-shell references, an SCF stability analysis (with an optional descent along the unstable direction) can be performed. Fractional occupations (Fermi–Dirac, Gaussian, Methfessel–Paxton [71], and Marzari–Vanderbilt [72] smearing) on the Mermin free-energy surface [73] are available for metallic and other difficult systems, including gradients, geometry optimization, and semi-numerical frequencies.

SCF convergence is reached when three criteria are satisfied simultaneously in two consecutive iterations: the change of the total energy $|\Delta E| < 10^{-8}$ Ha, the maximum change of the density matrix $\max|\Delta P| < 10^{-6}$, and the norm of the DIIS residual $\| \mathbf{FPS} - \mathbf{SPF} \| < 10^{-5}$ a.u. All convergence thresholds can be changed by the user.

### 3.8 Geometry optimization, transition states, and reaction paths

Geometry optimization in StudDFT is performed by default with the trust-region rational function optimization (TR-RFO) method in redundant internal coordinates; for diatomic molecules, and at the user's request, the BFGS method in Cartesian coordinates is used, with Powell damping to preserve positive definiteness [74]. The initial Hessian is the diagonal Lindh model Hessian [75] (default) or the full coupled Almlöf–Lindh model. In TR-RFO, the step length is limited by an adaptive trust radius with an initial value of 0.15 Bohr. The radius is increased by a factor of 1.2 when the actual and predicted energy changes agree well (ratio >0.75), decreased by a factor of 2 for poor agreement (ratio <0.25), and by a factor of 4 if the ratio is negative. The optimization terminates successfully when four conditions are met simultaneously: the maximum gradient component $<4.5\cdot10^{-4}$ Ha/Bohr, the RMS gradient $<3.0\cdot10^{-4}$ Ha/Bohr, the maximum atomic displacement $<1.8\cdot10^{-3}$ Bohr, and the RMS displacement $<1.2\cdot10^{-3}$ Bohr. The optional keyword `%opt_tight` makes all four thresholds 30 times tighter (they then coincide with the "tight" criteria of Gaussian), and `%opt_verytight` sets the maximum-force threshold to $2\cdot10^{-6}$ Ha/Bohr, the value of the "very tight" criteria of Gaussian.

The redundant internal coordinates [76] are formed from the connectivity determined by the covalent radii [77] and include (1) the stretches of all atom pairs closer than $1.3(R_a + R_b)$, where $R_a$ and $R_b$ are the covalent radii of the atoms; (2) the valence angles of all unique triples of bonded atoms A–B–C; and (3) the dihedral angles of all unique quartets of bonded atoms A–B–C–D. Since version 1.29.0, weakly bound complexes, whose covalent graph consists of several disconnected fragments (hydrogen-bonded, halogen-bonded, and van der Waals complexes, ion pairs), are treated by adding auxiliary interfragment bonds: all interfragment contacts shorter than 0.85 of the sum of the van der Waals radii [46], plus the shortest interfragment distances needed to connect all fragments. The angles and torsions across these bonds are generated automatically, and floors are imposed on their model force constants to keep the steps bounded. Previously, such systems were optimized in Cartesian coordinates, which converged slowly along the soft intermolecular directions. The cutoff of the singular values when forming the redundant-coordinate transformation is $10^{-6}$.

In redundant internal coordinates, individual bonds, angles, and dihedral angles can be frozen, which makes partial (constrained) optimizations possible. Rigid and relaxed PES scans are performed by specifying the geometry as a Z-matrix together with the scanned coordinates, steps, and numbers of points. Transition-state searches use the partitioned rational function optimization (P-RFO) method [78, 79] in redundant internal coordinates: along the selected mode (the reaction coordinate) a step is taken toward the energy maximum, and along the remaining modes toward the minimum; the followed mode is identified by the maximum overlap with that of the previous step. The Hessian is updated by the Bofill formula [80] and, optionally, recomputed exactly every $N$ steps. The initial Hessian (and, optionally, the geometry) is read from a file written by a previous frequency calculation, or computed at the starting geometry. A frequency calculation can follow the location of the transition state automatically. Intrinsic reaction coordinate (IRC) paths are followed from the transition state in both directions with the Gonzalez–Schlegel method [81]. Born–Oppenheimer molecular dynamics on the ground-state surface (velocity-Verlet, NVE) is also available.

### 3.9 Hessian, vibrational frequencies, and harmonic analysis

The Hessian is computed analytically for HF and DFT (including ECPs, closed-shell range-separated hybrids, and closed-shell RI-J with pure functionals) and semi-numerically, i.e., by central differences of the analytical gradients, for MP2, UMP2, and PUMP2 (for PUMP2, the UMP2 Hessian is used), PCM, smearing, open-shell range-separated hybrids, and the remaining RI cases; the D3 dispersion contribution is also obtained by finite differences of its gradient. The semi-numerical route can be requested for any method (for verification). For the meta-GGA functional $r^2$SCAN, the analytical Hessian of the distributed executables (which are built without libxc) uses an approximate built-in kernel, which does not contain the $\tau$-dependent terms for open shells; its frequencies deviate from the semi-numerical ones by up to 68 $cm^{-1}$ for closed shells (the torsion of $C_2H_6$; 24 $cm^{-1}$ even for $H_2$) and by up to 264 $cm^{-1}$ for open shells ($CH_3$) in the test set of Table 3, so that the semi-numerical route (`%hess_method numerical`) must be used for $r^2$SCAN frequencies (for BLYP, the two routes agree within 0.1 $cm^{-1}$). The finite-difference step is $10^{-3}$ Bohr by default and can be changed with the keyword `%hess_step`. In the semi-numerical route, only the symmetry-unique atoms are displaced, and the remaining rows of the Hessian are reconstructed using the point group of the molecule. The MP2 and UMP2 Hessians are an exception: they use a separate driver that displaces all atoms with a fixed step of $5\cdot10^{-3}$ Bohr.

The analytical Hessian is one of the most complex parts of the program. Its most expensive step is the solution of the CPKS equations, in which the exchange-correlation kernel is integrated numerically. Since version 1.28.0, this kernel is integrated on a separate, compact response grid (grid 11 by default, keyword `%cpks_grid`), which is a valid quadrature in its own right. For example, for $(H_2O)_2$ at the B3LYP/6-311++G(2d,2p) level, the default response grid (34 306 points) reproduces all vibrational frequencies obtained with the full production grid (135 605 points) to 0.03 $cm^{-1}$. Every analytical Hessian is accompanied by a check of its translational and rotational invariance at the stationary point (the ratios $\max|\mathbf{HT}|/|\mathbf{T}|$ and $\max|\mathbf{HR}|/|\mathbf{R}|$, which vanish for an exact Hessian), which detects quadrature artifacts and unconverged geometries.

After computing the Cartesian Hessian $H_{xy}$, the harmonic frequencies are obtained by the standard procedure: mass weighting $H'_{xy} = H_{xy}/\sqrt{m_x m_y}$, projection onto the space orthogonal to the six (five for linear molecules) external degrees of freedom, and diagonalization of the projected matrix. IR intensities are computed from the analytical dipole derivatives (on the analytical-Hessian route). The zero-point vibrational energy and the thermodynamic functions are computed in the rigid-rotor–harmonic-oscillator approximation; imaginary frequencies are marked with a special flag. For closed shells, the static dipole polarizability can be obtained from the CPKS equations for an external field together with the analytical frequencies.

### 3.10 User interface

The StudDFT input language is deliberately simple and follows the “one parameter, one line” principle, with keywords beginning with the % sign. The geometry is given in Cartesian coordinates or as a Z-matrix (since version 1.30.2, the format of a `%geo` block is recognized automatically) and is terminated by a blank line or the end of the file. An example of an input file for the geometry optimization and frequency calculation of $H_2O$ at the B3LYP/6-31G(d,p) level is given in Table S2 of the Supplementary Information. It is important for students that the program never terminates with cryptic diagnostics upon a user error. Therefore, error situations when reading the input (a nonexistent keyword or basis set, incompatible charge and multiplicity, etc.) are explicitly checked and reported through a unified error handler.

# 4 Program capabilities

Since version 1.14.0, StudDFT has been distributed in two editions built from the same source code. The Student Edition contains all features except three advanced subsystems (constrained DFT, real-time TDDFT, and Ehrenfest dynamics), which are available only in the Full Edition; all other keywords and numerical results are identical in the two editions. The capabilities of version 1.30.2 are summarized below; the version in which a feature appeared is given in parentheses where it is later than 0.82.

- *Electronic structure methods*: RHF/UHF; RKS/UKS with LDA (SVWN, SVWN3), GGA (BLYP, BP86, BPW91, PBE), meta-GGA ($r^2$SCAN; present since v0.79, verified in this work), global hybrid (B3LYP, B3LYP3, PBE0), and range-separated hybrid ($\omega$B97, $\omega$B97X, $\omega$B97X-D; v0.88) functionals; RMP2/UMP2/PUMP2 with frozen core.
- *Job types*: single-point energies; analytical gradients; geometry optimization with frozen coordinates; harmonic frequencies, IR intensities, and thermochemistry; transition-state search (P-RFO) with optional frequency verification; IRC (v0.86); rigid and relaxed PES scans; Boys–Bernardi counterpoise correction [82]; Born–Oppenheimer molecular dynamics (v1.11).
- *Excited states*: linear-response TDDFT/TDHF and TDA for closed-shell references and all functionals except meta-GGA (v0.87), singlet and triplet excitations, oscillator strengths, transition and unrelaxed state dipole moments with generalized Mulliken–Hush couplings [83], export of excitation amplitudes for Multiwfn and TheoDORE.
- *Solvation and dispersion*: IEFPCM, C-PCM, and COSMO with 18 tabulated solvents or a user-defined permittivity, analytical PCM gradients, semi-numerical PCM frequencies (v0.93), LR-PCM for TDDFT (v0.94); D3(BJ) dispersion [84, 85] (energies, gradients, and finite-difference Hessians; native implementation since v1.13) and the Chai–Head-Gordon dispersion of $\omega$B97X-D.
- *Performance options*: RI-J and RI-JK density fitting (v0.96–1.02); direct and incremental SCF; optional multigrid SCF; OpenMP parallelization of all time-critical steps (one-electron integrals, Fock builds, XC quadrature, gradients, CPHF/CPKS, TDDFT); symmetry-blocked numerical Hessians.
- *SCF control*: core, SAD, and checkpoint (v1.08) guesses; DIIS, damping, level shifting, MOM; fractional-occupation smearing with fixed or relaxed spin moment (v1.07–1.09); stability analysis of closed-shell solutions, except for meta-GGA functionals (v1.24).
- *Analysis*: Mulliken charges and dipole moment; Löwdin, Hirshfeld [86], and CM5 [87] charges; Mayer [88], Mulliken, and Löwdin bond orders and valences; spin populations (v1.05); point-group detection with irreducible-representation labels of the MOs and normal modes (v1.06–1.09); static dipole polarizability and finite external electric field (v1.04).
- *Full Edition only*: charge-constrained DFT [89] on Becke fragments (v0.84), real-time TDDFT with several propagators [90] (v1.10), and Ehrenfest dynamics [91] including smeared references (v1.11–1.12).

After a Hessian calculation (`%job freq` or `opt+freq`), the program writes a `*.molden` file containing the geometry, the energy, the vibrational frequencies, and the normal-mode vectors in the format of the Molden program [92]. This widely used format allows the geometry to be visualized and the normal vibrations to be animated with Molden, ChemCraft [93], or Avogadro [94]. The geometries of all optimization cycles in the StudDFT output files are also recognized by the Moltran program [95] (command: `moltran <file.out> /nxyz0@%geo`). Molecular orbitals can be saved in the Molden format, which is understood by Molden, ChemCraft, and Multiwfn [96]. The Hessian is also written to a `*.hess` file, which can be used to start a transition-state search or an IRC calculation.

# 5 Verification

In this section, the results of StudDFT are compared with those of reference quantum chemistry packages on a test set of molecules, in order to demonstrate that the program reproduces the reference data in all its operating modes within the accuracy expected for the corresponding algorithms.

All StudDFT calculations were carried out with the default parameters, except for the choice of the integration grid, the use of spherical basis functions, and, in a few cases, additional SCF convergers. The reference programs were ORCA 6.0.1 [6], PySCF 2.14.0 [10] with the exchange-correlation library libxc 7.0.0 [32], and, in a few cases, NWChem 6.8 [7]. Identical basis sets from the Basis Set Exchange [50] were used (for the exceptions, see Section 5.3). A major problem in comparisons of DFT results is the choice of an integration grid analogous to that of the reference program. Different programs use different types of grids, numbers of nodes, and pruning schemes, which are often not fully documented, and frequently include additional, undocumented features intended to speed up the calculation. For the single-point energy and gradient tests, StudDFT grid 8 (Table S1) was used, which approximately corresponds to grid 5 of ORCA with `IntAcc=6` (for the meta-GGA r$^2$SCAN, however, grid 5 of ORCA is much less accurate; Section 5.1). In the optimization and Hessian tests, this ORCA grid does not permit the calculation of the Hessian; therefore, for those calculations, the ORCA grid `defgrid3` (`IntAcc=6`, `Z_IntAccXC=6`) was chosen. In all ORCA reference calculations, the RI and COSX approximations were switched off (keywords NoRI and NoCOSX; the only exception is the MP2 reference of $C_6H_6$, see Table 3, note c), so that the Coulomb and exchange terms are evaluated exactly in both programs.

The StudDFT results of Tables 1 and 3, of the Pople rows of Table 2, and of Table S3 were obtained with version 1.30.2 (Windows, 6 OpenMP threads), except for the PBE set and one BLYP calculation of Table 3 (version 1.29.8), the MP2 set of Table 3 (version 1.29.6), and the optimization times of Table S3 (versions 1.29.2–1.29.8); the semi-numerical r$^2$SCAN Hessians of Table 3 and the replacement calculations of Tables 1 and 2 (Table 1, notes a and c; $O_2$ and $CHF_2O$ with 6-311++G(2d,2p)) were carried out with a Linux build of version 1.30.2 (the Hessians at the geometries optimized under Windows). These versions contain all corrections discussed in Section 5.3. The changes between versions 1.29.2 and 1.30.2 (threading of the open-shell XC Hessian, zero-sized matrix blocks, the convergence threshold of the response equations, the dispersion correction in semi-numerical Hessians, diagnostics, the transition-state and IRC drivers, and the input parser) do not change the test results beyond the printed precision. For the BLYP optimization set, version 1.30.2 reproduces the results of version 1.29.2 within $3\cdot10^{-11}$ Ha in the energies, $2\cdot10^{-7}$ Å in the geometries, and 0.01 cm$^{-1}$ in the frequencies, and a Linux build of version 1.30.2 reproduces the BP86 set of version 1.29.5 (Windows) within 0.04 μHa, 0.01 cm$^{-1}$ on average (0.25 cm$^{-1}$ at most), and $7\cdot10^{-6}$ Å on average ($2\cdot10^{-4}$ Å at most, for a soft torsion). For the BP86 optimization set, versions 1.29.8 and 1.30.2 give identical results (energies within 0.001 μHa, frequencies within 0.01 cm$^{-1}$).

The test set consisted of 50 species representing characteristic classes of electronic structure: the atoms H, He, Li; the ions $Na^+$, $H_2^+$, $OH^-$; the homonuclear diatomic molecules $H_2$, $N_2$, $F_2$, $Cl_2$; the hydrides HF, HCl, $BeH_2$, $BH_3$, $NH_3$, $PH_3$, $SiH_4$; the halides NaCl, $AlF_3$, $SiF_4$; the oxides CO, $CO_2$, $SO_2$, $H_2O$, $O_3$; the hydrocarbons $CH_4$, $C_2H_2$, $C_2H_4$, $C_2H_6$, $C_6H_6$; the organic compounds HCN, $H_2CO$, HCOOH, $CH_3OH$, $CH_3COH$, $CH_3COOH$, $C_2H_5OH$, $CHF_2OH$, $C_6H_5OH$, $C_6H_5NH_2$, $C_4H_4NH$ (pyrrole); and the radicals, biradicals, and carbenes OH, NO, $CH_3$, $CH_3O$, $CHF_2O$, $O_2$ (triplet and singlet), and $CH_2$ (triplet and singlet). The initial geometries corresponded to typical experimental bond lengths and angles. The complete set of inputs and outputs is available in the program repository.

## 5.1 Single-point energies and gradients

Table 1 gives the mean deviations (MD) and the mean absolute deviations (MAD) of the energies and of the root-mean-square gradients for the $N$ = 50 test molecules (the definitions are given in the Supplementary Information). The MD of the energies computed by StudDFT at the same geometries agree with the reference programs to within 3 μHa, with the exception of the PBE and PBE0 functionals, for which the StudDFT energies are systematically higher (MD = MAD = 63.2 and 47.1 μHa relative to ORCA and 30.8 and 23.1 μHa relative to PySCF), and of the meta-GGA r$^2$SCAN, whose deviations from ORCA are caused by the quadrature (see below). The shifts of PBE and PBE0 are due mainly to the value of the constant $\mu$ of the PBE exchange. StudDFT uses the value $\mu = 0.21951$ given in the original publication [25], libxc [32] uses $\mu = \beta\pi^2/3 = 0.2195149727 ...$, and ORCA 6.0.1 uses $\mu = 0.21952$ (as printed in its output). Smaller contributions come from the constants of the PBE correlation (StudDFT: $\beta = 0.066725$ and $\gamma = 0.031091$ [25]; libxc: $\beta = 0.06672455060314922$ and $\gamma = (1 - \ln 2)/\pi^2$; ORCA prints $\beta = 0.066725$) and from the PW92 parameters $A$ of its local part (Section 2.3; StudDFT: the published values, libxc and ORCA: those of the reference code). To establish the origin of the shifts unambiguously, StudDFT 1.30.2 was rebuilt with these constants changed one at a time, and the energies were computed at the ORCA-optimized geometries of the 30 molecules of Table 3. The mean offset relative to ORCA (PBE: 50.1 μHa; PBE0: 37.2 μHa) consists of 51.9 μHa (PBE0: 38.9 μHa) due to the value of $\mu$, −3.3 μHa due to the PW92 parameters (this part also appears as the offsets of BP86 and BPW91 in Table 3), and 1.6 μHa due to the more precise value of $\gamma$ or $\beta$ (the energies do not distinguish between the two); with all three changes, the StudDFT energies agree with those of ORCA within 0.5 μHa on average (MAD; BLYP: 0.7 μHa). With identical quadratures, the PBE and PBE0 energies of StudDFT exceed those of PySCF by 25.6 and 19.2 μHa on average for these 30 molecules (30.8 and 23.1 μHa for the 50 molecules of Table 1), and a build with the libxc constants reproduces the PySCF energies within 0.01 μHa. The smaller shifts given in the journal version (29.9 and 21.0 μHa relative to ORCA and −1.8 and −2.2 μHa relative to PySCF) were obtained with a test build of version 0.79 in which $\mu$, $\beta$, and $\gamma$ had been replaced by the libxc values; this change was not carried over to the later versions (0.81 and later). We therefore attribute the deviations to the constant $\mu$ and, to a much smaller extent and with the opposite sign, to the local part of the correlation, which was assumed in the journal version to be the main cause. The agreement of the programs is much better for the gradients, which agree to within a few μHa/Bohr for HF and the LDA, GGA, and hybrid functionals (MAD at most 5.2 μHa/Bohr relative to ORCA and 0.2 μHa/Bohr relative to PySCF). For MP2, the energies agree within 0.7 μHa (with the same frozen core; Table 1, note c), whereas the RMS gradients deviate by 25.3 μHa/Bohr on average, by up to 424 μHa/Bohr for NO. PySCF with the same frozen core reproduces the ORCA gradients within 3.1 μHa/Bohr (RMS), whereas the RMS gradients of StudDFT deviate from those of PySCF by 6–56 μHa/Bohr for $H_2O$, CO, and $N_2$ and by 110–421 μHa/Bohr for the open-shell molecules $CH_3$, $O_2$, and NO; without the frozen core, StudDFT and PySCF agree within 0.03 μHa/Bohr (2.8 μHa/Bohr for NO). We attribute this deviation to the frozen-core MP2 gradient of StudDFT: finite-difference checks with version 1.30.2 show that the frozen-core gradient components differ from the derivative of the frozen-core MP2 energy by 10 and 15 μHa/Bohr ($H_2O$ and $SO_2$, cc-pVDZ), whereas the all-electron MP2 gradients agree with the finite differences within about 1 μHa/Bohr.

The largest energy deviations between the programs occur for molecules with several F atoms ($AlF_3$, $SiF_4$) (Table 1). This reflects the large density gradients near these nuclei, which require an increased grid density in this region. A comparison with a much denser grid (PySCF, 250×1202 points per atom, which agrees with 150×590 points within 3 μHa) shows that these deviations are quadrature errors, mainly of grid 8: with SVWN, the energies obtained with grid 8 are too high by 45.7 and 51.1 μHa for $AlF_3$ and $SiF_4$ (14.5 μHa for $CHF_2OH$), whereas the errors of the ORCA energies amount to −30.9 and 11.8 μHa with grid 5 and do not exceed 3 μHa with `defgrid3`. For PBE and PBE0, the lists of the largest deviations also reflect the offset due to the constants, which grows with the number of electrons (hence $Cl_2$). The meta-GGA r$^2$SCAN is much more sensitive to the quadrature: relative to ORCA, its energies deviate by 150.5 μHa on average (MAD; up to −573.5 μHa for $SO_2$) and its RMS gradients by 31.7 μHa/Bohr. For $SO_2$, $Cl_2$, $SiF_4$, $O_3$, $AlF_3$, and $H_2O$, the quadrature errors of the ORCA grid 5 amount to 110–594 μHa in absolute value, those of grid 8 to 1–142 μHa (the largest for $AlF_3$ and $SiF_4$), so that these deviations come mainly from the ORCA grid. For the RMS gradients of $C_2H_2$ (the largest deviation from ORCA, 230.5 μHa/Bohr) and $Cl_2$, the quadrature errors amount to 46 and 27 μHa/Bohr with grid 8 and to −185 and −18 μHa/Bohr with the ORCA grid 5; accurate r$^2$SCAN gradients thus require a denser grid than grid 8 as well. With the identical quadrature, StudDFT reproduces the PySCF r$^2$SCAN energies within 1.8 μHa on average (MD −1.6 μHa, due to the PW92 parameters $A$; Section 2.3) and the RMS gradients within 0.5 μHa/Bohr; the largest gradient deviations (5.6 and 5.2 μHa/Bohr for OH and $OH^-$) reflect small errors of the analytical r$^2$SCAN gradient of StudDFT: for the closed-shell molecules $OH^-$, HF, $H_2O$, and $H_2$, it differs from the finite differences of the StudDFT energy by 9.0, 3.7, 1.3, and −0.6 μHa/Bohr, respectively, whereas the PySCF gradients agree with their finite differences within 0.01 μHa/Bohr. In the results published for versions 0.79–0.81, NaCl also showed large deviations (43–194 μHa, depending on the method), and $BeH_2$ showed a deviation of −25 μHa at the MP2 level. A large part of these deviations (43–107 μHa for NaCl and −23 μHa for $BeH_2$ with MP2) was caused by the basis-set library: the $d$-polarization exponents of the cc-pVDZ sets of Li, Be, Na, and Mg distributed with these versions were taken from a preliminary version of these sets, whereas the BSE contains the published sets of Prascher et al. [97] (for Na, the exponent of the $d$ function is 0.0928 instead of 0.0973; for Be, 0.2354 instead of 0.238; all other exponents and contraction coefficients coincide). The library was updated in version 0.81.5, and the values in Table 1 take this update into account; for all other molecules checked (Section 5.3), the single-point results of version 1.30.2 coincide with those of the earlier versions. With HF (for $CHF_2O$, the ground state; see below), the energies agree with ORCA within 0.63 μHa ($C_6H_5OH$) and the RMS gradients within 3.2 μHa/Bohr ($CH_3O$: at its $C_{3v}$ starting geometry, the two programs converge to different solutions of the degenerate $^2$E state, $C_s$-symmetric in ORCA and unsymmetric in StudDFT, which have nearly the same energy but differ in the forces on the H atoms).

**Table 1.** Mean deviations (MD) and mean absolute deviations (MAD) of the total energies (μHa) and of the RMS gradients (μHa/Bohr) computed by StudDFT 1.30.2 relative to ORCA 6.0.1 and PySCF 2.14.0 (in square brackets) for single-point energy/gradient calculations of 50 test molecules with the cc-pVDZ basis set (StudDFT: grid 8; ORCA: grid 5 with `IntAcc=6` and the NoRI and NoCOSX keywords; PySCF: the StudDFT basis-set file and the same quadrature as StudDFT, Treutler–Ahlrichs M4 with 75×302 points per atom without pruning; the StudDFT and PySCF gradients include the response of the grid weights).

| | Energy | | | RMS gradient | |
|---|---|---|---|---|---|
| Theory | MD | MAD | Largest deviations (>10 μHa) | MD | MAD |
| HF[a] | 0.0 | 0.0 | – | 0.1 | 0.1 |
| SVWN | 2.5 | 3.9 | 77.6 ($AlF_3$), 39.2 ($SiF_4$), 16.5 ($CHF_2OH$) | 0.4 | 3.9 |
| BLYP | 2.8 | 3.9 | 69.8 ($AlF_3$), 62.3 ($SiF_4$), 12.8 ($CHF_2OH$) | −0.3 | 3.1 |
| BP86 | −2.5 | 6.7 | 55.4 ($AlF_3$), 36.0 ($SiF_4$), −16.1 ($C_6H_5NH_2$) | −0.8 | 3.8 |
| BPW91 | −1.0 | 5.3 | 54.1 ($AlF_3$), 37.9 ($SiF_4$), −10.4 ($SO_2$) | −0.7 | 3.8 |
| PBE | 63.2 | 63.2 | 231.5 ($SiF_4$), 208.0 ($AlF_3$), 149.2 ($Cl_2$) | −1.5 | 5.2 |
| | [30.8] | [30.8] | [95.2 ($SiF_4$), 77.5 ($AlF_3$), 75.8 ($Cl_2$)] | [0.0] | [0.1] |
| PBE0 | 47.1 | 47.1 | 169.9 ($SiF_4$), 155.0 ($AlF_3$), 111.0 ($Cl_2$) | −0.2 | 3.6 |
| | [23.1] | [23.1] | [71.2 ($SiF_4$), 58.2 ($AlF_3$), 56.6 ($Cl_2$)] | [0.1] | [0.2] |
| B3LYP | 2.2 | 3.2 | 57.2 ($AlF_3$), 49.0 ($SiF_4$), 10.8 ($CHF_2OH$) | −0.4 | 2.6 |
| r$^2$SCAN[b] | −113.1 | 150.5 | −573.5 ($SO_2$), 414.1 ($Cl_2$), −376.5 ($SiF_4$) | 2.2 | 31.7 |
| | [−1.6] | [1.8] | [–] | [−0.1] | [0.5] |
| MP2[c] | 0.0 | 0.1 | – | 1.7 | 25.3 |

[a] For the $CHF_2O$ radical, the UHF ground state obtained with `%initial_guess core` is used; with the default settings, StudDFT converges to a solution 3.3 mHa higher, whereas ORCA gives the ground state. [b] The deviations from ORCA are due mainly to the ORCA quadrature: with grid 5, ORCA 6.0.1 uses a Gauss–Chebyshev radial grid, Lebedev grids with at most 434 points, pruning, and Becke weights (about 14 000–15 000 points per non-hydrogen atom, against 22 650

points per atom for StudDFT grid 8), whose errors for this meta-GGA reach several hundred μHa (Section 5.1). [c] For Li, $BeH_2$, $Na^+$, and NaCl, the frozen core of ORCA (no frozen orbitals for Li and Be, the 1s orbital of Na) was specified with `%frozen_core`; the default frozen core of StudDFT, the noble-gas core of every atom, contains in addition the 1s orbital of Li and Be and the 2s and 2p orbitals of Na and raises the MP2 energies of these four species by 0.19–1.08 mHa.

The stability of the StudDFT results under variation of the basis type and size is demonstrated in Table 2 for HF calculations. The StudDFT results agree well with the reference program for both the energies and the gradients, with polarization and diffuse functions and with ECPs. The energy deviations of 2.7 μHa for cc-pVTZ are due entirely to two molecules, $BeH_2$ (48.4 μHa) and NaCl (83.9 μHa), which also account for most of the gradient deviation (1.4 of 1.5 μHa/Bohr): unlike the cc-pVDZ file, the cc-pVTZ file distributed with StudDFT still contains the preliminary *d* and *f* exponents of Li, Be, Na, and Mg, whereas the BSE library used by ORCA contains the published sets of Prascher et al. [97] (the same applies to the cc-pVQZ and aug-cc-pV*n*Z files of version 1.30.2; all other elements of these files coincide with the BSE). With identical basis sets, the HF/cc-pVTZ energies of StudDFT and PySCF coincide within 0.26 μHa for all 19 molecules checked.

The open-shell UHF calculations require attention to the electronic state. The UHF ground states were identified with PySCF stability analysis. For triplet $O_2$, the default SCF settings of version 1.30.2 (SAD guess, MOM during the first iterations) give the ground state with the cc-pVDZ and cc-pVTZ basis sets, but an excited solution with the 6-31G(d,p) and 6-311++G(2d,2p) basis sets (e.g., −148.938695 Ha instead of −149.611927 Ha with 6-31G(d,p)); with the strict Aufbau occupation (`%uks_occupation strict`), the ground state is obtained with the Pople basis sets but not with the correlation-consistent ones (with an additional level shift, `%level_shift 0.2`, the strict occupation again leads to an excited solution with 6-311++G(2d,2p), 164 mHa above the ground state). For the $CHF_2O$ radical, the default settings lead to a solution about 3 mHa above the ground state with the 6-31G(d,p) and cc-pVDZ basis sets, whereas the core-Hamiltonian guess gives the ground state; with cc-pVTZ, the situation is reversed, and with 6-311++G(2d,2p), two runs with identical input and default settings converged to different solutions (in the tests of Table 2, a Windows run with 6 threads gave the solution 2.9 mHa above the ground state and a Linux run with 2 threads the ground state, which is used in Table 2). In the cc-pVDZ single-point test (Table 1), ORCA converged to the ground state of $CHF_2O$, whereas StudDFT with the default settings gave the solution 3.3 mHa higher. In the comparisons, the same (ground) state must therefore be ensured in both programs. With DFT, the default and the strict settings give the same results for all open-shell molecules tested.

**Table 2.** Mean deviations (MD) and mean absolute deviations (MAD) of the energies (μHa) and of the RMS gradients (μHa/Bohr) computed by StudDFT relative to ORCA (in square brackets: relative to PySCF 2.14) for single-point HF energy/gradient calculations of the test molecules as a function of the basis set (6-31G(d,p): 50 molecules; 6-311++G(2d,2p): 49 molecules, since this basis set is not defined for He in the BSE).

| | Energy | | RMS gradient | |
|---|---|---|---|---|
| Basis set | MD | MAD | MD | MAD |
| 6-31G(d,p) | 0.0 | 0.0 | 0.0 | 0.1 |
| 6-311++G(2d,2p) | 0.0 | 0.0 | 0.0 | 0.1 |
| cc-pVTZ[a] | 2.7 | 2.7 | 1.5 | 1.5 |
| LANL2DZ+ECP, analytical integrals[b] | 8.1 | 13.8 | −15.5 | 22.4 |
| | [7.8] | [14.1] | [−16.1] | [22.9] |
| LANL2DZ+ECP, numerical integrals[b] | 0.4 | 4.2 | 5.0 | 5.7 |
| | [0.1] | [0.2] | [4.5] | [5.3] |

[a] The deviations are due to the Be and Na basis sets (see the text). [b] 15 molecules containing the elements Na–Ca, Ti, Br, and I ($AlF_3$, $Br_2$, $Ca_2$, $Cl_2$, HBr, HCl, HI, $I_2$, KCl, NaCl, $PH_3$, $SO_2$, $SiF_4$, $SiH_4$, $TiCl_4$).

The numerical and analytical ECP methods give nearly identical results (within about 0.5 μHa) for most molecules, but the analytical integrals deviate from PySCF by −47, +53, and +112 μHa for $AlF_3$, HI, and $I_2$, respectively (and from the numerical method by −46, +51, and +112 μHa). With the numerical method, the StudDFT energies agree with PySCF within about 2 μHa (MAD 0.25 μHa). The remaining gradient deviations, which are similar relative to ORCA and PySCF, are largest for $AlF_3$ and HI, whose largest gradient components differ from PySCF by 93 and 45 μHa/Bohr in the numerical method (for $AlF_3$, the PySCF gradient was confirmed by finite differences of the energy); for the other molecules, they do not exceed 5 μHa/Bohr. Finite differences of the StudDFT energy show that these deviations arise in the ECP gradient itself: for $AlF_3$ and HI, the gradient of the numerical method differs from the derivative of the StudDFT energy by 76 and 36 μHa/Bohr. The cause is the radial integration interval of the ECP gradient integrals, which lacks the extension to the off-centre maximum of the integrand introduced for the energy integrals in version 0.77.1. With this extension, the gradients of $AlF_3$ and HI agree with the finite differences of the energy within 0.1 μHa/Bohr; this correction is not yet included in version 1.30.2.

## 5.2 Geometry optimization and vibrational frequencies

Table 3 demonstrates the agreement between StudDFT and ORCA for geometry optimizations and vibrational frequencies at the located stationary points. These results depend on several algorithms at once: the energies and gradients, the construction of internal coordinates, the optimization algorithm, and the Hessian. A reduced set of 30 molecules was used (listed in Section S4 of the Supplementary Information). The energies, all interatomic distances in the optimized structures, and all vibrational frequencies were analyzed simultaneously. With the RI and COSX approximations switched off in ORCA, the energies of the optimized structures agree within 0.7 μHa on average (MAD) for HF, SVWN, BLYP, and B3LYP (the largest deviation, −8.8 μHa, is that of $C_6H_6$ with BLYP); the offsets of BP86 and BPW91 (−4.0 and −3.6 μHa) and of PBE and PBE0 (50.1 and 37.2 μHa) are caused by the different constants of these functionals, mainly by the constant $\mu$ of the PBE exchange (Section 5.1). The energy offsets of 89–144 μHa in the optimization tests of the journal version [15], attributed there to the denser ORCA grid, were in fact caused by the RI approximation, which had remained switched on in those ORCA runs (RI-J and RIJCOSX by default): for $BeH_2$ (B3LYP), the offset of 22.9 μHa nearly vanished without RI (0.02 μHa), whereas a change of the ORCA grid (defgrid2 instead of defgrid3) changes the energy by only 0.24 μHa. The interatomic distances and vibrational frequencies agree well: the mean absolute deviations of the interatomic distances do not exceed $1.2\cdot10^{-4}$ Å, and those of the vibrational frequencies amount to 0.19–0.25 $cm^{-1}$ for HF and the LDA, GGA, and hybrid functionals, 0.9 $cm^{-1}$ for MP2, and 1.1 $cm^{-1}$ for $r^2$SCAN (Table 3). The largest individual deviations are those of the torsional mode of $CH_3OH$ (2.3 $cm^{-1}$ with HF and 2.1 $cm^{-1}$ with SVWN) and of the stretching mode of the OH radical (up to 2.0 $cm^{-1}$ with BLYP, PBE0, and B3LYP); with $r^2$SCAN, the largest deviation is that of the umbrella mode of $CH_3$ (5.6 $cm^{-1}$). On average, the frequencies of StudDFT are 0.09–0.16 $cm^{-1}$ higher than those of ORCA with HF and the LDA, GGA, and hybrid functionals and 1.0 $cm^{-1}$ with $r^2$SCAN. In the results published for versions 0.79–0.81, deviations of the frequencies of 10–15 $cm^{-1}$ were observed for $O_3$, $CH_4$, and $SiH_4$ at several DFT levels; as shown in Section 5.3, they were caused by defects of the CPKS implementation and are absent in version 1.30.2. Note that the frequency statistics are computed over the 162 genuine vibrations of the 30 molecules; the journal version averaged over all Cartesian degrees of freedom, including the translations and rotations (166 of the 330 values for the molecule set used there), whose values are zero in both programs, which lowered the published frequency MD and MAD by a factor of about two. The interatomic distances were computed from the full-precision Cartesian coordinates of the outputs. The $r^2$SCAN functional, which was present in version 0.82 but not yet sufficiently tested and was therefore not described in the journal version, is included in Table 3 with the semi-numerical Hessian (the analytical $r^2$SCAN Hessian of version 1.30.2 is not yet accurate enough; see Section 3.9). With identical quadratures, the $r^2$SCAN energies of StudDFT are lower than those of PySCF (libxc) by 1.4 μHa on average; with the PW92 parameters of the reference code, which libxc uses in $r^2$SCAN as well, they agree within 0.01 μHa, except for the OH and NO radicals (1.0 and 0.5 μHa). The larger scatter of the $r^2$SCAN energies relative to ORCA (MAD 11.0 μHa) is the same for PySCF (MD 0.2, MAD 10.9 μHa relative to ORCA at the ORCA geometries) and is therefore not caused by the implementation of the functional in StudDFT (ORCA 6.0.1 evaluates r$^2$SCAN through libxc, version 6.2.2); it reflects the quadrature errors of both grids for this meta-GGA (at the geometries of Table 1, those of grid 8 amount to 25 and 17 μHa for $SO_2$ and $O_3$; Section 5.1).

**Table 3.** Mean deviations (MD) and mean absolute deviations (MAD) of the total energies, interatomic distances, and vibrational frequencies for 30 molecules optimized by StudDFT 1.30.2 (PBE: version 1.29.8) at various levels of theory with the cc-pVDZ basis set, relative to ORCA 6.0.1 (DFT calculations with the grid `defgrid3`, `IntAcc=6`, `Z_IntAccXC=6.0`, and the NoRI and NoCOSX keywords). The frequency statistics refer to the genuine vibrations.

| | Total energy, μHa ($N = 30$) | | Distances, $10^{-3}$ Å($N = 213$) | | Frequencies, cm$^{-1}$ ($N = 162$) | |
|---|---|---|---|---|---|---|
| Theory | MD[a] | MAD[a] | MD | MAD | MD | MAD |
| HF | 0.0 | 0.0 | 0.00 | 0.08 | 0.09 | 0.24 |
| SVWN | −0.1 | 0.7 | 0.03 | 0.09 | 0.09 | 0.19 |
| BLYP | −0.3 | 0.7 | 0.01 | 0.03 | 0.14 | 0.19 |
| BP86 | −4.0 | 4.0 | 0.04 | 0.08 | 0.13 | 0.20 |
| BPW91 | −3.6 | 3.6 | 0.04 | 0.08 | 0.11 | 0.19 |
| PBE | 50.1 | 50.1 | 0.03 | 0.07 | 0.13 | 0.19 |
| PBE0 | 37.2 | 37.2 | −0.02 | 0.09 | 0.16 | 0.23 |
| B3LYP | −0.2 | 0.6 | 0.05 | 0.12 | 0.09 | 0.25 |
| r$^2$SCAN[b] | −1.2 | 11.0 | 0.08 | 0.11 | 1.04 | 1.14 |
| MP2[c] | 0.5 | 0.6 | 0.04 | 0.12 | 0.00 | 0.91 |

[a] The deviations for BP86, BPW91, PBE, and PBE0 are caused by different constants of these functionals in the two programs: the PW92 parameters of the local correlation (−3 to −4 μHa) and, for PBE and PBE0, mainly the constant $\mu$ of the PBE exchange (0.21951 in StudDFT and 0.21952 in ORCA). With $\mu = 0.21952$, the PW92 parameters of ORCA, and the exact $\gamma$ (or $\beta$), StudDFT reproduces the ORCA energies within 0.5 μHa (MAD; BP86 and BPW91: 0.6 μHa; Section 5.1). [b] Semi-numerical Hessian (finite differences of the analytical gradients). [c] Earlier test set (version 1.29.6; HCOH instead of HCN): 26 molecules, 164 distances, and 123 vibrations (ORCA MP2 results for $BeH_2$, $C_2H_2$, $C_2H_4$, and $C_2H_6$ were not available); the energy statistics exclude $C_6H_6$, for which the ORCA reference was computed with RI-MP2 (with $C_6H_6$: MD −3.6, MAD 4.7 μHa).

## 5.3 Corrections after version 0.82 and their effect on the results

The following changes in the program and its data affect the results reported in the journal version [15] (obtained with versions 0.79–0.81). They are listed here together with their effect, because version 0.82 is still distributed for compatibility with the journal article.

1. *CPKS exchange-correlation kernel* (versions 0.52.7–1.27). To save memory and time, the kernel of the CPKS equations was evaluated on a regular subsample (every *n*-th point) of the production grid. A subset of a Lebedev shell is not a quadrature: the response part of the Hessian lost its rotational invariance and acquired spurious components of about $5\cdot10^{-3}$ Ha/Bohr$^2$ along the rotations. For isolated molecules, these components are largely removed by the projection of the rotations, but the vibrational frequencies are distorted by up to 17 cm$^{-1}$, and the degenerate modes are split (by up to 19 cm$^{-1}$ in the test set); for weakly bound complexes, whose librations are hindered rotations, imaginary frequencies could appear. The subsampling was active only above a certain problem size; among the 31 molecules of the optimization test inputs (the journal version used 30 of them, the ORCA reference for $H_2^+$ having failed; Table 3 of this work uses HCN instead of the duplicate HCOH), it affected $C_2H_2$, $C_2H_4$, $C_2H_6$, $C_6H_6$, $CH_3OH$, $CH_4$, $CO_2$, $H_2CO$, HCOH, HCOOH, $O_3$, $SO_2$, and $SiH_4$. Since version 1.28.0, the kernel is integrated on a separate response grid that is a valid quadrature (Section 3.9), and every Hessian is accompanied by an invariance check.

2. *P86 kernel* (corrected in version 1.26.0). In the closed-shell (RKS) route, the second-derivative kernel of the P86 correlation used the PZ81 local part, whereas the energy and potential used PW92. The analytical BP86 Hessian therefore differentiated a slightly different functional; the effect is largest for soft modes in regions of low density. For the molecules of the test set not affected by item 1, the closed-shell BP86 frequencies changed by 0.1–2.1 cm$^{-1}$, whereas the open-shell (UKS) frequencies did not change (≤0.02 cm$^{-1}$).

3. *Radial scale of the Mura–Knowles grids 1–5* (corrected in version 1.28.0, for the elements of group 2 in version 1.29.2; see Section 3.6). The verification calculations of this work, performed with grid 8, are not affected, but default-grid DFT energies of earlier versions may contain quadrature errors of up to several mHa.

4. *Basis-set library*: the *d* exponents of the cc-pVDZ sets of Li, Be, Na, and Mg (updated to the BSE values in version 0.81.5; the cc-pVTZ, cc-pVQZ, and aug-cc-pV*n*Z files still contain the preliminary sets of these four elements).

5. *Missing second derivative of the Becke weights* in the analytical Hessian (identified in version 1.28.2, not yet included; see Section 2.2).

In addition, the interfragment coordinates of the optimizer (version 1.29.0) do not affect the test set, all of whose molecules are covalently connected, and the tighter optimization criteria (version 1.29.5) are optional. Three further defects were identified in the course of this work and are not yet corrected in version 1.30.2: the radial integration interval of the ECP gradient integrals (Section 5.1; up to 76 μHa/Bohr for $AlF_3$ in the numerical method), the approximate frozen-core term of the MP2 gradient (10–15 μHa/Bohr per component for $H_2O$ and $SO_2$; RMS deviations from PySCF of up to 56 μHa/Bohr for closed shells and 421 μHa/Bohr for the open-shell NO; Section 5.1), and small errors of the analytical r$^2$SCAN gradient (up to 9 μHa/Bohr for $OH^-$; Section 5.1). All three affect only the gradients (and the geometries and frequencies derived from them) and not the energies. In addition, the analytical r$^2$SCAN Hessian of the distributed executables is approximate (Section 3.9).

**Table 4.** Mean absolute deviations (maximum absolute deviations in parentheses) of the harmonic frequencies (cm$^{-1}$) computed by StudDFT versions 0.79–0.81 and 1.30.2 from the PySCF 2.14 analytical frequencies at the same geometry (optimized with version 1.30.2), with the same basis set (cc-pVDZ) and the same quadrature (Treutler–Ahlrichs M4, 75×302 points per atom, no pruning). For BP86, the PySCF reference uses the PW92-based P86, as StudDFT.

| | $CH_4$ | | $O_3$ | | $SiH_4$[a] | |
|---|---|---|---|---|---|---|
| Functional | 0.79–0.81 | 1.30.2 | 0.79–0.81 | 1.30.2 | 0.79–0.81 | 1.30.2 |
| SVWN | 4.1 (13.6) | 0.08 (0.17) | 7.8 (9.3) | 0.17 (0.30) | 1.7 (8.4) | 0.53 (1.4) |
| BLYP | 5.5 (14.4) | 0.07 (0.15) | 12.2 (17.2) | 0.53 (0.65) | 3.1 (13.3) | 0.92 (2.4) |
| BP86 | 5.9 (13.7) | 0.07 (0.14) | 8.9 (13.0) | 0.53 (0.68) | 3.0 (7.9) | 1.68 (4.3) |
| BPW91 | 6.6 (12.3) | 0.11 (0.20) | 7.1 (14.9) | 0.78 (1.06) | 3.1 (8.4) | 1.26 (3.3) |
| PBE | 3.8 (9.8) | 0.08 (0.18) | 9.9 (15.1) | 0.53 (0.71) | 2.9 (6.3) | 1.68 (4.3) |
| PBE0 | 3.5 (8.1) | 0.08 (0.16) | 8.3 (13.9) | 0.27 (0.41) | 2.3 (5.1) | 1.32 (3.4) |
| B3LYP | 3.6 (12.0) | 0.07 (0.18) | 9.3 (15.0) | 0.38 (0.48) | 2.0 (3.8) | 0.61 (1.6) |

[a] The remaining differences for $SiH_4$ are due to the grid-weight response, whose first-derivative terms are included in the StudDFT Hessian but neglected in the PySCF analytical Hessian (verified for BLYP; see the text).

To verify the corrected Hessian independently of ORCA, the harmonic frequencies of the three molecules with the largest published deviations ($CH_4$, $O_3$, and $SiH_4$) were recomputed with PySCF 2.14 at the StudDFT-optimized geometries, using the same basis set and an identical quadrature (the StudDFT basis-set file and the Treutler–Ahlrichs grid with 75 radial and 302 angular points per atom). With identical quadratures, the total energies of StudDFT and PySCF agree within 0.01 μHa for SVWN, BLYP, BP86 (with the PW92-based P86), and B3LYP, and within 0.4 μHa for BPW91; the PBE and PBE0 energies differ by 8–41 μHa because of the different constants of the PBE functional, mainly $\mu$ (first-order estimates with $\mu = 0.21951$ reproduce these differences within about 0.3 μHa; a build of StudDFT with the libxc constants reproduces the PySCF energies within 0.01 μHa; see Section 5.1). The frequency deviations are given in Table 4. For $CH_4$ and $O_3$, the mean absolute deviations of version 1.30.2 from PySCF do not exceed 0.11 and 0.78 cm$^{-1}$ (maximum deviations 0.20 and 1.06 cm$^{-1}$), respectively, whereas the earlier versions deviated by up to 17 cm$^{-1}$. The triply degenerate modes of $CH_4$ and $SiH_4$, which were split by up to 19 cm$^{-1}$ in versions 0.79–0.81, are degenerate within 0.1 cm$^{-1}$ ($CH_4$) and 0.2 cm$^{-1}$ ($SiH_4$) in version 1.30.2. For $SiH_4$, the bending frequencies of StudDFT differ from the PySCF analytical values by −1.4 to +4.3 cm$^{-1}$, depending on the functional. For BLYP, it was verified that this difference is due to the response of the grid weights: the frequencies obtained by finite differences of the PySCF gradients that include the grid-weight response (BLYP: 880.7, 880.8, 881.6, 947.7, and 948.2 cm$^{-1}$) agree with the StudDFT analytical values (880.7, 880.7, 880.9, 947.7, and 947.8 cm$^{-1}$) and with the StudDFT semi-numerical Hessian (880.8, 880.8, 880.9, 947.8, and 947.8 cm$^{-1}$), but not with the PySCF analytical Hessian, which neglects the grid response (878.3, 878.7, 878.9, 947.4, and 947.8 cm$^{-1}$). Recomputation of the CPKS kernel on the full production grid (`%cpks_grid same`) changes the StudDFT frequencies by at most 0.03 cm$^{-1}$.

The single-point energies and gradients of version 1.30.2 were also compared with those of the earlier versions on identical inputs. Apart from the effects of the basis-set update (Be, Na) and the states of $O_2$ and $CHF_2O$ discussed above, they coincide: the HF/cc-pVDZ energies and gradients of the 49 molecules common to the single-point sets of versions 1.29.2 and 1.30.2 (Table 1) are identical (within $10^{-12}$ Ha); for HF with the 6-31G(d,p) (50 molecules), 6-311++G(2d,2p) (48 molecules), and cc-pVTZ (7 molecules) basis sets and with the LANL2DZ ECP (15 molecules, numerical and analytical integrals), the energies agree within 0.004 μHa for closed shells and within 0.2 μHa for the open-shell radicals, and the DFT energies and gradients at the starting geometries of the optimization tests ($CH_4$, $O_3$, and $SiH_4$ with all functionals and all 31 molecules of the BP86 inputs) agree to all printed digits, except for $BeH_2$. The optimized geometries agree within $7\cdot10^{-5}$ Å, except for $CH_3OH$ ($2\cdot10^{-4}$ Å) and $C_2H_6$ (BP86, $9\cdot10^{-4}$ Å), for which the optimizer followed slightly different paths along the soft torsions (the final energies differ by less than 0.3 μHa).

### 5.4 Performance

Table S3 of the Supplementary Information shows typical computation times of StudDFT for different types of calculations, and Fig. S1 shows the speedup due to parallelization. The calculations were carried out on a personal computer with a 6-core Intel Core i5-10600K processor (4.1 GHz), 16 GB RAM, and Windows 10; the program was built with Intel Fortran with speed optimization, and 6 OpenMP threads were used. The times in Table S3 were measured in series of consecutive runs and are therefore somewhat conservative: with pauses between the runs, the 6-thread optimization and frequency calculation of pyrrole takes about 25% less time (Section S3). Although the run time varies strongly with the system size and the method, a DFT optimization with a frequency calculation takes less than 2 minutes for most test molecules and about 10 minutes for benzene, which makes the program quite suitable for practical classes in quantum chemistry and molecular modeling.

## 6 Discussion

The verification makes it possible to derive the following quantitative characteristics. For the single-point HF, MP2, and DFT energies, the mean absolute deviation of the StudDFT results from the reference programs is no worse than 7 μHa (0.004 kcal/mol) for basis sets of the cc-pVDZ type (except for PBE and PBE0, see below, and for r$^2$SCAN relative to ORCA, whose deviations are due to the quadrature; Section 5.1), which is far below the typical errors in the energetics of chemical reactions. This deviation is determined mainly by the residual differences in the grids and convergence parameters; with identical basis sets and quadratures, StudDFT and PySCF agree within 0.01 μHa for most functionals. Systematic differences arise only from different conventions: the constants of the PBE functional (mainly the constant $\mu$ of the PBE exchange; several tens of μHa for PBE and PBE0), the PW92 parameters of the local correlation (3–4 μHa relative to ORCA), the local part of P86 (relative to Gaussian and libxc), the versions of basis sets in different libraries, and the default frozen core of MP2 for Li, Be, and Na (Table 1, note c). For the optimized geometries, the MAD of the interatomic distances does not exceed $1.2\cdot10^{-4}$ Å, which is more than sufficient for practical applications. For the harmonic vibrational frequencies, the MAD over the entire set does not exceed 0.25 $cm^{-1}$ for HF and the LDA, GGA, and hybrid functionals, 0.91 $cm^{-1}$ for MP2, and 1.14 $cm^{-1}$ for $r^2$SCAN. The consistency of the numerical and analytical results confirms the correctness of the analytical gradients and Hessians, except the analytical $r^2$SCAN Hessian of the distributed executables (Section 3.9), the frozen-core MP2 gradient (Section 5.1), and, to a small extent, the analytical $r^2$SCAN gradient (deviations from the finite differences of up to 9 μHa/Bohr, for $OH^-$; Section 5.1). Apart from the frozen-core UMP2 gradients, the results for closed-shell and open-shell systems are characterized by similar error magnitudes. Calculations with various basis sets and with ECPs also agree with the reference ones.

The history of the defects listed in Section 5.3 is instructive for the development of quantum chemistry codes, in particular with the help of artificial intelligence tools. The defect of the CPKS kernel did not affect the energies (for $(H_2O)_2$, the total energy changed by about $10^{-7}$ Ha, while the lowest frequency changed by more than 200 $cm^{-1}$, from an imaginary to a real value), the analytical-versus-numerical Hessian check in the test suite used a basis set too small to activate the subsampling, and for isolated molecules a large part of the error is removed by the projection of the rotations. Such defects are therefore revealed only by comparisons that include soft modes, weakly bound complexes, and degenerate vibrations, and by permanent internal diagnostics such as the invariance check of the Hessian.

At present, the program has a number of limitations, some of which are planned to be removed in the course of further work. Minnesota-type meta-GGA and double hybrid functionals are not implemented. TDDFT is available only for closed-shell references, not for meta-GGA functionals, and only for single-point calculations (no excited-state gradients); the analytical Hessian is not available for MP2, PCM, smearing, open-shell range-separated hybrids, and some RI cases (the semi-numerical route is used instead), and the second derivative of the grid weights is not yet included in the analytical Hessian. For open-shell UHF calculations, the SCF may converge to an excited solution depending on the initial guess and the occupation control (for triplet $O_2$, the default settings fail with the Pople basis sets and the strict Aufbau settings with the correlation-consistent ones), so that a comparison of the energies obtained with different settings is recommended in difficult cases; an automatic stability analysis of unrestricted solutions is not yet available. The ECP gradients of the numerical method and the frozen-core MP2 gradients contain errors of up to several tens of μHa/Bohr (several hundred μHa/Bohr for open-shell UMP2), the analytical $r^2$SCAN gradient errors of up to 9 μHa/Bohr (Sections 5.1 and 5.3), and $r^2$SCAN frequencies require the semi-numerical Hessian. The MP2 gradient keeps several four-index arrays of the dimension of the basis set in memory (3 GB each for 140 basis functions), one of them per OpenMP thread: with 6 threads, the MP2 gradient of $C_6H_5NH_2$/cc-pVDZ needs about 25 GB, more than the 16 GB of RAM of the computer used here, so that it runs with paging and is slower than on a single thread (Table S3, note c). Molecular symmetry is detected and used for the analysis of orbitals and normal modes and for reducing the number of displacements in semi-numerical Hessians (except for MP2), but not for reducing the cost of the SCF. Only the LANL2DZ ECPs are built in. Parallelization is limited to shared-memory (OpenMP) systems.

Performance optimization remains an important direction. Although much attention has been paid to this issue, there is considerable room for improvement, in particular in the Hessian calculation, in reducing the number of SCF cycles, and in optimizing the integration grids.

# 7 Conclusions

This work presents the quantum chemistry package StudDFT (version 1.30.2), a compact program for calculations of small and medium-sized molecular systems that implements the Hartree–Fock method, density functional theory with LDA, GGA, meta-GGA, global and range-separated hybrid functionals, and MP2 perturbation theory with spin-projected variants. The program supports energy calculations, geometry optimization, transition-state and reaction-path searches, harmonic vibrational analysis, effective core potentials, continuum solvation, dispersion corrections, density fitting, and linear-response TDDFT. Analytical first derivatives with respect to the nuclear coordinates are implemented for all ground-state methods except PUMP2 and constrained DFT, and analytical second derivatives for HF and DFT (for $r^2$SCAN, the semi-numerical Hessian is recommended; Section 3.9). The program was developed primarily for educational use. Its results have been systematically verified on a set of 50 molecules (30 in the optimization and frequency tests) by comparison with the reference packages ORCA and PySCF: single-point total energies are reproduced with an accuracy better than 7 μHa (apart from the systematic shifts of several tens of μHa for PBE and PBE0 caused mainly by a different value of the constant $\mu$ of the PBE exchange, and from the quadrature-related deviations of $r^2$SCAN from ORCA), interatomic distances in the optimized structures better than $1.2 \cdot 10^{-4}$ Å, and harmonic frequencies better than 1.2 $cm^{-1}$ (0.25 $cm^{-1}$ for HF and the LDA, GGA, and hybrid functionals; mean absolute deviations). The internal consistency of the analytical and numerical derivatives has also been confirmed. The program is a compact monolithic code and is characterized by ease of installation and learning and by low hardware requirements. We see StudDFT not as an alternative to existing "production-level" quantum chemistry packages such as Gaussian or ORCA, but as a complement to the ecosystem of educational software for quantum chemistry. We hope that StudDFT will prove useful both in quantum chemistry courses and in educational research projects.

## Code and data availability

The executables of StudDFT 1.30.2 (Student and Full Editions for Windows; Full Edition for Linux), the short manual, the test inputs and outputs, and the version 0.82.1 corresponding to the journal article are available at https://github.com/skignatov/StudDFT-release. The binaries are distributed under the StudDFT Binary Freeware License, which permits free use for educational, academic, research, and personal purposes (commercial use requires permission of the author) and redistribution of the complete, unmodified distribution archive; the source code is not distributed at present. The D3 reference data and the numerical libraries included in the distribution are covered by their own licenses (LGPL-3.0 for the D3 data files).

## Acknowledgments

Parts of the program code and of the text of this manuscript were prepared with the assistance of the large language model Claude (Anthropic). The author designed the program, verified the code and all results, and takes full responsibility for the content of this paper.

# Supplementary Information

## S1. Predefined integration grids

**Table S1.** Predefined DFT integration grids of StudDFT 1.30.2 (keyword `%grid` with the grid number; grids 6–10 can also be selected by the names `fine`, `ultrafine`, `tm`, `g09`, and `orca7`, and grids 11 and 12 as response grids by `%cpks_grid light` and `tight`). The radial grids are those of Mura and Knowles (Log3) [64], Euler–Maclaurin (Murray, Handy, and Laming) [65], and Treutler–Ahlrichs [66]; the angular grids are Lebedev–Laikov grids [67].

| Grid | Name | Radial grid / points | Lebedev points (pruning zones) | $\chi$ adjustment |
|---|---|---|---|---|
| 1 | coarse | Mura–Knowles / 25 | 110 | none |
| 2 | medium | Mura–Knowles / 50 | 110 / 302 | none |
| 3 | standard[a] | Mura–Knowles / 75 | 110 / 302 / 590 | none |
| 4 | extended | Mura–Knowles / 100 | 110 / 302 / 590 | none |
| 5 | extreme | Mura–Knowles / 150 | 110 / 302 / 590 | none |
| 6 | fine | Euler–Maclaurin / 75 (H, He: 39) | 50 / 86 / 266 / 302 / 266 | Becke |
| 7 | ultrafine | Euler–Maclaurin / 99 (H, He: 51) | 50 / 86 / 434 / 590 / 434 | Becke |
| 8 | tm | Treutler–Ahlrichs M4 / 75 | 302 | Treutler |
| 9 | g09 | Euler–Maclaurin / 75 | 302 | Becke |
| 10 | orca7 | Treutler–Ahlrichs M3 / 39–64[b] | 110 / 434 / 590 / 770 / 590 | none |
| 11 | response_light[c] | Euler–Maclaurin / 35 | 110 / 302 | Becke |
| 12 | response_tight | Euler–Maclaurin / 50 | 110 / 302 | Becke |

[a] Default production grid. Since version 1.28.0, the radial scale of grids 1–5 is that of Mura and Knowles (see the main text). [b] Row-dependent number of radial points (Krack–Köster scheme [98]), as in ORCA; for H and He, the angular orders are one tier lower (50 / 302 / 434 / 590 / 434). [c] Default grid for the exchange-correlation kernel of the CPKS equations (keyword `%cpks_grid`). Three-tier pruning (grids 2–5, 11, and 12): 110 points for $r < 0.35\,R$ and $r > 8\,R$, 302 points for $0.35\,R < r < R$ and $4.5\,R < r < 8\,R$, and the largest order for $R < r < 4.5\,R$, where $R$ is the Bragg–Slater radius; the five-zone schemes of grids 6 and 7 follow Gill, Johnson, and Pople [99].

## S2. Example input file

**Table S2.** A typical input file of StudDFT (geometry optimization and frequency calculation of the water molecule at the B3LYP/6-31G(d,p) level).

```
! Water full optimization and frequencies -- Comment on any line
%method b3lyp       ! In-line comments are also possible
%basis  6-31G(d,p)  ! Basis set from the open library
%job    opt+freq    ! Optimization followed by frequency calculation
%charge 0           ! Charge of the molecule (default 0)
%mult   1           ! Spin multiplicity (default 1)
%geo                ! Start of Cartesian coordinates (end: blank line)
O1   0.000000   0.000000   0.117790
H2   0.000000   0.755453  -0.471161
H3   0.000000  -0.755453  -0.471161
```

## S3. Computation times and parallel speedup

**Table S3.** Computation times for the test sets of molecules (6 OpenMP threads, Intel Core i5-10600K, Windows 10; single points: version 1.30.2, optimizations: versions 1.29.2–1.29.8).

| | 1 SCF + gradient, 50 molecules | | Optimization + frequencies, 30 molecules | | |
|---|---|---|---|---|---|
| Theory | Total time, s | Per molecule, s | Optimization cycles | Total time, s | Per molecule, s |
| HF | 136 | 3 | 165 | 536 | 18 |
| SVWN | 171 | 3 | 168 | 787 | 26 |
| BLYP | 256 | 5 | 167 | 1512 | 50 |
| BP86 | 273 | 5 | 166 | 1485 | 50 |
| BPW91 | 249 | 5 | 169 | 1419 | 47 |
| PBE | 269 | 5 | 173 | 1590 | 53 |
| PBE0 | 252 | 5 | 156 | 1304 | 43 |

| Theory | 1 SCF + gradient, 50 molecules | | Optimization + frequencies, 30 molecules | | |
|---|---|---|---|---|---|
| | Total time, s | Per molecule, s | Optimization cycles | Total time, s | Per molecule, s |
| B3LYP | 258 | 5 | 167 | 1640 | 55 |
| $r^2$SCAN[a] | 416 | 8 | 160 | 1679 | 56 |
| MP2 | 3298[c] | 66 | 168 | 32955[b] | 1098 |

[a] Semi-numerical Hessian. [b] Including 29 318 s for $C_6H_6$, whose semi-numerical MP2 Hessian requires 72 MP2 gradients; the other 29 molecules take 3637 s. [c] Including 1218 s for $C_6H_5NH_2$, computed with one OpenMP thread: with 6 threads, its MP2 gradient needs about 25 GB of memory (Section 6).

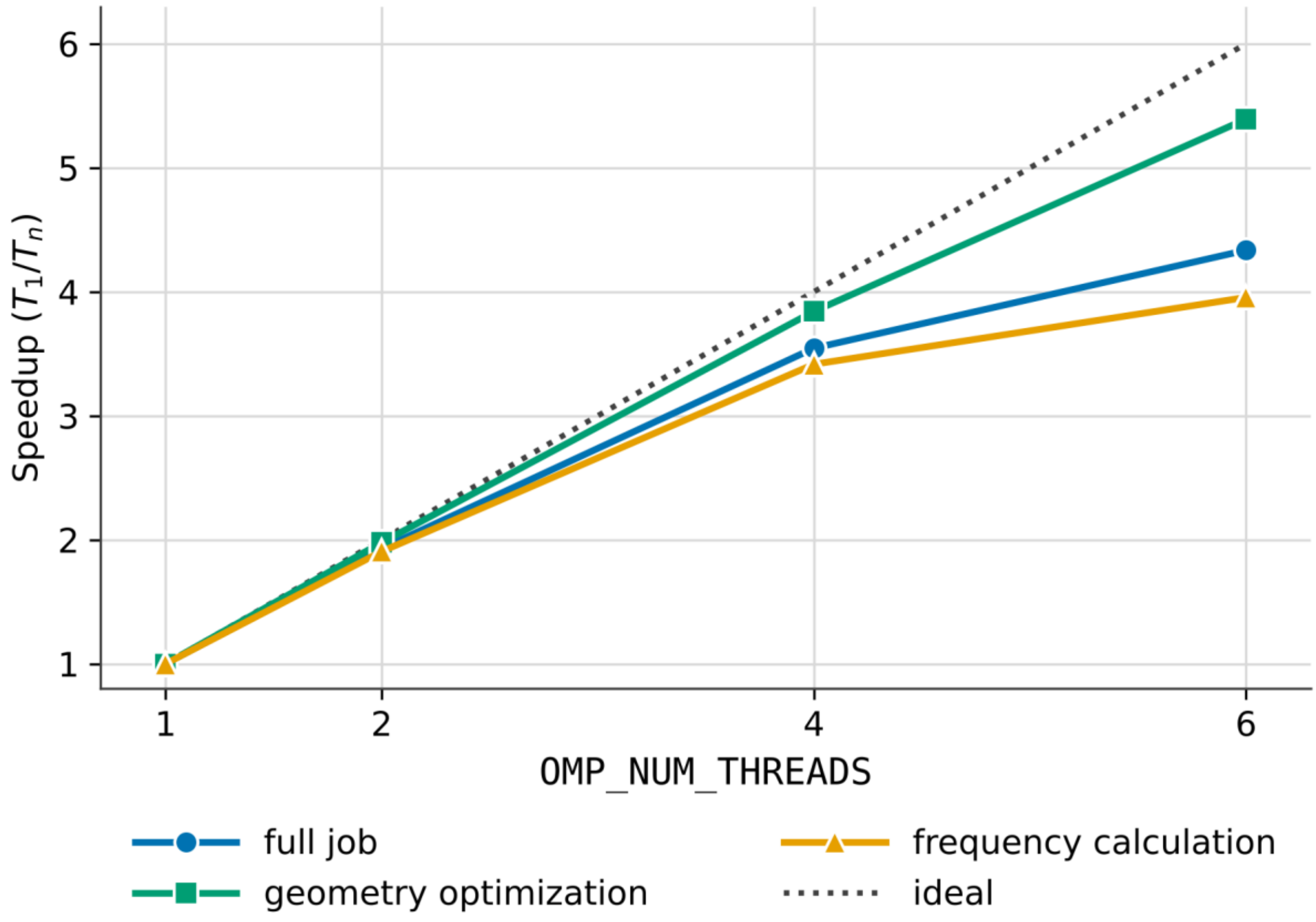

**Figure S1.** Speedup due to parallelization (opt+freq calculation of pyrrole, $C_4H_5N$, at the PBE0/cc-pVDZ level; version 1.30.2).

Figure S1 shows the speedup with the number of OpenMP threads; on 6 cores, it amounts to 5.4 for the geometry optimization, 4.3 for the full job, and 4.0 for the frequency calculation. The runs were started at intervals of several minutes. The two-electron integrals and their derivatives scale almost linearly (speedup 5.6 for the two-electron part of the Hessian), whereas the exchange-correlation contributions to the derivative Fock matrices and to the CPKS equations, which are accumulated point by point on the grid, scale worse (speedup 3.6 for the coupled-perturbed step) and limit the speedup of the frequency calculation. In a series of runs without pauses, the 6-thread job took 187 s instead of 137 s, presumably because the processor lowers its clock frequency under sustained load on all cores.

## S4. Molecules of the optimization tests

The optimization and frequency tests (Table 3) include the following 30 molecules: $OH^-$, OH, $Cl_2$, $H_2$, HF, HCl, CO, $N_2$, $F_2$, NO, $O_2$ (triplet), $BeH_2$, $H_2O$, $CH_2$ (singlet and triplet), $CO_2$, $O_3$, $SO_2$, $CH_3$, $CH_4$, $NH_3$, $SiH_4$, $H_2CO$, HCN, $C_2H_2$, $C_2H_4$, $CH_3OH$, HCOOH, $C_2H_6$, and $C_6H_6$. They contain 109 atoms, 213 interatomic distances, and 162 vibrational degrees of freedom. The MP2 test (Table 3, note c) used the earlier list, which contained HCOH (a second formaldehyde entry; see the main text) instead of HCN.

## S5. Definitions of MD and MAD

The mean deviation (MD) and the mean absolute deviation (MAD) used in the tables of the main text are

$$\mathrm{MD} = \frac{1}{N}\sum_{i=1}^{N}\left(X_i^{\mathrm{S}} - X_i^{0}\right), \qquad \mathrm{MAD} = \frac{1}{N}\sum_{i=1}^{N}|X_i^{\mathrm{S}} - X_i^{0}|,$$

where $X_i^{\mathrm{S}}$ and $X_i^{0}$ are the energy (or the RMS gradient, interatomic distance, or vibrational frequency) of item *i* computed with StudDFT and with the reference program, respectively, and *N* is the number of items.

## S6. Software implementation

StudDFT is written in standard Fortran 2003 and uses a number of language features of the Fortran 2008 standard. Fortran 2018 features and compiler-specific extensions are deliberately avoided, so that the program can be built without modification by GFortran (version 9 or newer) and by the Intel Fortran compilers (Classic and oneAPI ifx). Parallelism is implemented with OpenMP. Support for MPI and hybrid MPI+OpenMP schemes has deliberately been left outside the scope of the program: MPI requires the installation of additional libraries, which is often impossible for inexperienced users, and for the target class of problems (up to about 500 basis functions) OpenMP provides sufficient performance.

The Windows executables of version 1.30.2 are built with Intel Fortran and are monolithic files of 32 MB (Full Edition) and 22 MB (Student Edition); parallel execution additionally requires the Intel OpenMP library `libiomp5md.dll` (1.9 MB), which is included in the distribution. The Linux executable (Full Edition, 33 MB) is built with GFortran 11 and statically linked with OpenBLAS. The source code of the Full Edition of version 1.30.2 comprises 69 Fortran modules and the main program with a total of 109 thousand lines, including 77 thousand lines of code, 25 thousand lines of comments, and 7 thousand blank lines (version 0.82: 54 modules, 72 thousand lines). The program and the test infrastructure (inputs, outputs, and extraction scripts for all test calculations of this work) are available in the program repository.